\documentclass[journal=jacsat,manuscript=article]{achemso}
\usepackage{longtable}
\usepackage{amsmath}
\usepackage{subfigure}
\usepackage{multirow}
\usepackage{dcolumn}
\usepackage{xfrac}
\usepackage{arydshln}
\newcolumntype{d}[1]{D{.}{.}{#1}}
\usepackage[normalem]{ulem}
\usepackage[usenames,dvipsnames]{color}
\SectionNumbersOn
\usepackage{hyperref}
\usepackage{xr}

\newcommand*{\sups}[1]{$^{#1}$}
\newcommand*{\bfit}[1]{\textbf{\textit{#1}}}
\newcommand*{\epp}[0]{ESPResSo++}
\newcommand*{\lj}[0]{Lennard-Jones}
\newcommand*{\lebc}[0]{Lees-Edwards boundary condition}

\newcommand*{\gcell}[1]{\textbf{#1}$^{\prime}$}
\newcommand*{\lcell}[1]{\textbf{#1}$^{\prime}$$_\text{L}$}
\newcommand*{\rcell}[1]{\textbf{#1}$^{\prime}$$_\text{R}$}
\newcommand*{\rg}[0]{\textbf{r2g}}
\newcommand*{\gr}[0]{\textbf{g2r}}

\author{Zhen-Hao Xu}
\affiliation{Center for Data Processing, Johannes Gutenberg-Universität Mainz, Anselm-Franz-von-Bentzel-Weg 12, 55128 Mainz, Germany}
\author{James Vance}
\affiliation{Center for Data Processing, Johannes Gutenberg-Universität Mainz, Anselm-Franz-von-Bentzel-Weg 12, 55128 Mainz, Germany}
\author{Nikita Tretyakov}
\affiliation{Center for Data Processing, Johannes Gutenberg-Universität Mainz, Anselm-Franz-von-Bentzel-Weg 12, 55128 Mainz, Germany}
\author{Torsten Stuehn}
\affiliation{Max Planck Institute for Polymer Research,	Ackermannweg 10,
	55128 Mainz, Germany}
\author{Andr\'e Brinkmann}
\affiliation{Center for Data Processing, Johannes Gutenberg-Universität Mainz, Anselm-Franz-von-Bentzel-Weg 12, 55128 Mainz, Germany}
\email{zhexu@uni-mainz.de}
\date{\today}

\title{Implementation and Parallel Optimization of the Lees-Edwards Boundary Condition in ESPResSo++}

\begin{document}

\begin{abstract}

The Lees–Edwards boundary condition (LEbc) provides the possibility to simulate molecular or coarse grained systems under non-equilibrium conditions, namely by introducing a shear flow which has potential applications in high-speed fluids, thermoplastic and other non-equilibrium processes. This paper discusses a new LEbc implementation in the molecular dynamics (MD) software package \epp that focuses on the parallel efficiency of LEbc for scale-out simulations. Using the LEbc code, shear flow simulations were carried out for model systems such as Lennard-Jones fluids and Kremer-Grest polymer melts. Some important physical properties and phenomena, including the linear profiles of shear velocities, non-layered density distribution and shear thinning, have been successfully reproduced or captured. The results are also in good agreement with those from previous literature\cite{LJ-Ruiz-Franco2018} even with unphysical simulation conditions, which gives a solid validation to our implementation of LEbc. We considered in depth the parallelization of the LEbc code to efficiently scale in high performance computing (HPC) environments. A modified scheme for data communication has been introduced within the domain decomposition framework inside \epp. The presented benchmarks show a linear and good scaling for LEbc simulations for up to $1,024$ processor cores in supercomputer systems.
\end{abstract}

\clearpage

\section{Introduction}
Molecular dynamics (MD) describes the motion of a molecular system based on classical Newtonian mechanics, where particles or coarse grained beads are characterized by their masses and charges and where their dynamics is driven by empirical force fields\cite{MM-Leach2001,CHARMMFF}. An empirical force field usually approximates quantum mechanical calculations\cite{CGenFF, MMPT-Xu2019} or experimental observations\cite{MMPT-Oxa-XuMeuwly2017,MMPT-Mackeprang-2016}. The computer simulations provided by MD are an important and efficient tool to observe the motion of particles, compute physical quantities and compare with experimental observations (i.e., viscometric properties of fluids)\cite{Book_Understanding2002,Book_EVANS1990_NE_LIQUIDS}.  Non-equilibrium molecular dynamics (NEMD), as a variant of molecular dynamics, is required if the equilibrium conditions are not satisfied for a modeled system. Similar to its equilibrium counterpart, NEMD is based on time-reversible equations of motion. However, it differs from conventional mechanics by using a microscopic environment for the macroscopic second law of thermodynamics\cite{NEMD-Hoover2004}.
\\
\\
In recent decades, there has been a continuously increasing interest in using NEMD techniques to study behaviors of liquids under various types of flow fields\cite{NEMD-Sarman1998}. With these NEMD methods, one can accurately measure transport properties of fluids with molecule-like structures (e.g., realistic molecules or coarse grained beads)\cite{SLLOD_Todd2007}. A typical model system of fluids for which NEMD techniques are intensively employed is a shear flow system. In fluid mechanics, the concept of a shear flow corresponds to a type of fluid flow which is caused by external forces and driving adjacent layers of fluid to move parallel to each other and with different speeds. Meanwhile, viscous fluids resist against such shear motion.
\\
\\
For computer simulations, there have been a number of methods which enable the simulation of a shear flow. One of the simplest ways is to introduce moving walls at the boundaries of the shear plane \cite{KharedPY97}.
With the moving walls, a shear flow can be driven passively, while a downside of this method is that the periodicity upon the shear plane is thus closed. Moreover, the shear or Couette flows under the closed moving walls will introduce strong boundary and finite size effects, such as layering effects for sheared particles close to the boundaries\cite{LEBC-Rastogi1996-Wall}. An interesting method is the SLLOD  technique, which introduces the flow velocities into the equation of particle motions\cite{SLLOD_Tuckerman1997,SLLOD_Petravic1998,SLLOD_Todd2007,SLLOD-Daivis2006}.
With the SLLOD equation of motions, the thermostat only acts on the \emph{peculiar} velocities of particles without the contribution from shears. Other methods such as hard reflecting walls\cite{DLMESO-Seaton2013} and quaternion-based dissipative particle dynamics (QDPD)\cite{QDPD-Sims2004} were additionally developed in the past years.
\\
\\
To simulate the steady shear flow in a fully open-boundary space, the Lees-Edwards boundary condition (LEbc) was introduced by Lees and Edwards\cite{LEBC_Lees1972} in 1972. This technique has been proven advantageous to accurately capture the non-linear behavior of particles in the shear. LEbc does not require external forces to drive the flow, i.e., by applying moving walls, which posit parallel to and move against each other. Thereby, the finite size effects are avoided and using LEbc recovers spatial homogeneity and bulk behaviors of particles for small systems in the simulation\cite{LEBC-Wagner2002-LB}, especially if limited computer power is considered. 
Parallel computing is of great importance to scale MD simulations and it is absolutely necessary to already include scalability considerations during software development. For the implementation of the Lees-Edwards boundary condition, one of the first parallel algorithms was proposed and discussed by Rastogi \emph{et al.} 25 years ago\cite{LEBC-Rastogi1996}. In their work, the data information of \emph{virtual} particles (namely the ghost particles near to the shear boundaries and being mapped to their corresponding image boxes) were pre-stored in forms of data arrays which are assigned to the virtual processors (or sub-domains in the modern domain decomposition). These \emph{virtual} particles are finally collected and force calculations are carried out for pairs including such \emph{virtual} particles. They showed the overall computing overhead under the parallelization to LEbc has been less than $10\%$ out of the total simulation run-time.
\\
\\
Nowadays, efficient domain decomposition techniques have become the standard technique to scale MD software packages to HPC environments with many processors working in parallel.
Recently, Bindgen \emph{et al.} have implemented LEbc in the ESPResSoMD software (\epp~ and ESPResSoMD have the same roots, but have developed into completely different software packages). They introduced a scheme called \emph{columnar} domain decomposition and modified the patterns from that in the standard domain decomposition of cell communication for cells in the adjacent layers to the boundaries of the shear planes\cite{LEBC-Bindgen2021}. By this, no pairs for short-range interactions are missed or doubly counted and the patterns of cell communication can be kept unchanged during the simulation. Besides the works discussed above, very few LEbc implementations considered their parallelization in depth.
The corresponding optimization of performance and validation for shear simulations with many-particle systems with more than one million particles using modern high performance computing (HPC) environments are rarely found.
\\
\\
In this work, we discuss the parallelization of the Lees-Edwards boundary condition by improving the communication scheme in the domain decomposition which greatly reduce the computing overhead and hence enhances the overall performance for HPC systems.
Our implementation of LEbc is also advantageous as it can be used in more generalized applications. Using this LEbc is not exclusive to specific potentials, MD types or thermostats.
The portability to open-source MD simulation packages is also easy to achieve if the scheme is properly adopted.
\\
\\
In this paper, we present the implementation of the Lees-Edwards boundary condition into the \epp
~MD simulation software\cite{EPP-Halverson2013,EPP-Guzman2019}.
In Section \ref{sec:method}, we first give an overview on how to model a steady shear flow under LEbc.
Second, we introduce the fundamental scheme for implementing LEbc by optimizing cell communication in the domain decomposition.
That helps capture all non-redundant particle-particle pairs for the force calculation to short-ranged based interactions (e.g., the \lj~interaction).
We also discuss the generalization from cell to node communication of LEbc and corresponding adaption to multi-core parallel computing systems.
In Section \ref{sec:result}, we present simulation experiments starting from the simple Lennard Jones fluid\cite{LJ-Ruiz-Franco2018} to million-particle Kremer-Grest polymer melts\cite{KG-Grest1986,KG-Kremer1990}.
We compare results with those from previous literatures and benchmark the parallelism performance in a super computing system.
In Section \ref{sec:summary}, we draw our conclusion and give the outlook on the future work.
\\
\\
The source code of \epp~with LEbc is available at (\url{https://github.com/xzhh/espressopp}) and the GPL-3.0 license is applied.
\section{Methods and Development}\label{sec:method}

\subsection{An Introduction to the Lees-Edwards boundary condition}\label{sec:lebc}

\begin{figure}[H]
	\begin{center}
		\includegraphics[trim=0px 0px 0px
		0px,clip,width=0.7\linewidth]{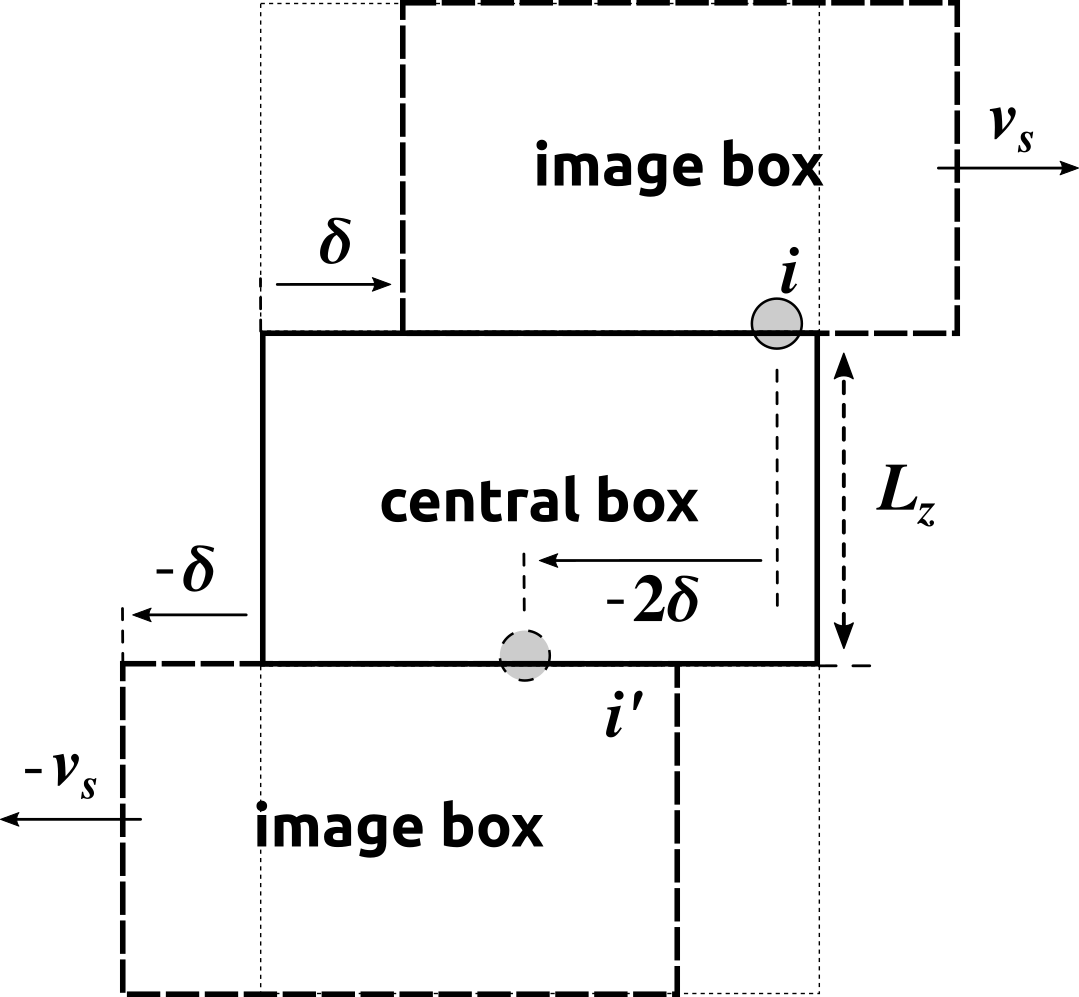}\\
		\caption{The schematic of the Lees-Edwards boundary condition. At a shear rate of $ \dot{\gamma} $, the image boxes move away from the central box at the speeds of $ v_s=\dot{\gamma}L_z/2 $ and $ -v_s $, respectively. At the time $ t $, the displacements are given by $ \delta=\dfrac{1}{2}\dot{\gamma} L_z t $ and $ -\delta $, compared to the original positions of the image boxes at $ t=0 $, see the dot-dashed frames. When the particle \emph{i} moves out of the top of the central box and re-enters at the bottom, then a shift of $ -2\delta $ is updated in the shear direction, and vice versa.}
		\label{fig:lebc}
	\end{center}
\end{figure}

Figure \ref{fig:lebc} depicts a typical schematic for implementing the Lees-Edwards boundary condition into a cuboid simulation system.
In addition to the Brownian motion, which is driven by conservative forces (such as the Lennard-Jones forces), the shear contribution for the motions of particles is applied systematically and presented in a form of the velocity with a linear profile, which writes as 
\begin{equation}
\begin{aligned} 
v_s(z)=\dot{\gamma}\cdot(z-\dfrac{L_z}{2}).
\end{aligned}
\label{eq:shear-rate}
\end{equation}
The shear speed $v_s$ is applied along the $x$-direction (the shear direction) and determined by the shear rate $ \dot{\gamma} $ and the distance of a particle from the $xy-$plane (the shear plane) at $ z=L_z/2 $ where $L_z$ is the height of the simulation box.
The $z-$direction is considered as the gradient direction and the third direction, $y$, is unrelated to shear flow.
This assignment of directions is fixed in all following discussion in this paper.
\\
\\
In LEbc, the periodicity is activated for all dimensions but the boundary cross for a particle is different when it occurs in the $xy-$planes of the boundaries (namely the top and bottom of the central box)
Here, we assume the particle $i$ is about to move out of the central box at a speed
\begin{equation}
\begin{aligned} 
\mathbf{p}_i(t)/m_i=\left[\left(v_x(t)+v_s(z(t))\right),v_y(t),v_z(t)\right],
\end{aligned}
\label{eq:vi}
\end{equation}
where $ \left[v_x(t),v_y(t),v_z(t)\right] $ is from kinematic contribution and is so-called \emph{peculiar} velocity\cite{DPD_SoddemannKremer2003,LEBC_Moshfegh2015,LJ-Ruiz-Franco2018},The position of $ i $ is written as
\begin{equation}
\begin{aligned} 
\mathbf{q}_i(t)=\big[x(t),y(t),z(t)\big].
\end{aligned}
\label{eq:qi}
\end{equation}
If the particle $i$ leaves from the top of the central box, the shear effect is taken into account as $ i $ is re-entering the central box (from the bottom) with its new position (see $ i' $ in Figure \ref{fig:lebc}), 

\begin{equation}
\mathbf{q}_i^\text{new}=\begin{bmatrix}
\mathbf{mod}\big(x(t)+\dot{\gamma} L_z t,L_x\big) \\
y(t) \\
z(t)-L_z
\end{bmatrix}.
\label{eq:pos_new}
\end{equation}
Here, \textbf{mod} takes a modulo of the new $ x $ position over $ L_x $, the length of the box along the $ x $-direction.
Changing the signs in Eq. \ref{eq:pos_new} can also describes a move from bottom to the top of the central box after a boundary cross.

\subsection{Shear Flow Simulation with LEbc}\label{sec:thermostat}

For a non-equilibrium simulations with a steady shear flow, the dynamics under the Langevin (LGV) thermostat can be presented as\cite{LJ-Ruiz-Franco2018,POLYM-Shang2017}

\begin{equation}
\begin{aligned} 
m_i\ddot{\mathbf{q}}_i(t)=\sum_{j\neq i}\mathbf{F}_{i,j}-\sum_{i} \xi(\mathbf{p}_i(t)-m_iv_s)+\sum_{i}\mathbf{F}^r_i.
\end{aligned}
\label{eq:langevin}
\end{equation}
\\
$ \mathbf{F}_{i,j} $ represents the conservative forces between particle $ i $ and other particles (here, we assume only a two-body system is considered).
$ (\mathbf{p}_i(t)-m_iv_s) $ represents the \emph{peculiar} velocity and $ \xi $ is the friction coefficient and this second term stands for the dissipative forces.
$ \mathbf{F}^r_i $ refers to the random forces.
Unlike an equilibrium MD (EQMD) simulation, the Langevin thermostat does not act on the absolute velocity of a particle but excludes the shear contribution.
Thus only the \emph{peculiar} part in Eq. \ref{eq:vi} participates in velocity integration in the \epp~software, combinedly with the force updates from the Langevin thermostat.
The shear contribution (into the simulations) can only take effects in the propagation of coordinates.
By this, the kinematic information along the $x-$direction is not lost and easy to access.
In order to validate the current LEbc development, we also aim to reproduce existing numerical experiments.\cite{LJ-Ruiz-Franco2018,POLYM-Shang2017}
Hence, the Langevin thermostat, in this work, is also altered to act on absolute velocities of particles for comparisons.
\\
\\
Despite of being well known and used in many simulation applications, the Langevin thermostat has an important drawback in which the total momentum of the system is not well conserved.
This is because the dragging force given by the thermostat is not pairwise.
The dissipative particle dynamics (DPD) method\cite{DPD-HoogerbruggeKoelman1992,DPD-Espanol1995,DPD-Groot1997,DPD_SoddemannKremer2003} is, in the contrary, a momentum conserving thermostat.
The dissipative forces in the DPD thermostat act on the relative velocities between two particles (or beads) and thereby pairwise (as well as random forces).
The DPD thermostat has a similar form as in Eq. \ref{eq:langevin}, but the dissipative and random forces are given by

\begin{equation}
\begin{aligned} 
\mathbf{F}^D_{i,j}=-\xi\left[\omega(r_{ij})\right]^2\left|\hat{r}_{ij}\cdot \vec{v}_{ij}\right|\cdot\hat{r}_{ij} 
\end{aligned}
\label{eq:dissipative}
\end{equation}
and
\begin{equation}
\begin{aligned} 
\mathbf{F}^R_{i,j}=\sqrt{24k_\text{B}T\xi}\cdot\theta_{ij}\omega(r_{ij})\left|\hat{r}_{ij}\cdot \vec{v}_{ij}\right|\cdot\hat{r}_{ij}.
\end{aligned}
\label{eq:random}
\end{equation}
where $ r_{ij}$ is the distance between particle \emph{i} and \emph{j}, $ \hat{r}_{ij}=\dfrac {\mathbf{q}_i-\mathbf{q}_j}{\left|\mathbf{q}_i-\mathbf{q}_j\right|}$ and $ \vec{v}_{ij}=v_i-v_j $.
$ \omega(r_{ij}) $ is the conditional weight function, which is expressed as
\begin{equation}
\omega(r_{ij}) = \left\{ \begin{array}{lr}
1-\dfrac{r_{ij}}{r_c}, & \text{if }r_{ij}<r_c, \\
0, & \text{if }  r_{ij}\geq r_c
\end{array}
\color{white}\right\}
,
\label{eq:omega}
\end{equation}
and $\xi$, $ k_\text{B} $, $ T $ and $ \theta_{ij} $ are respectively the friction coefficient, the Boltzmann constant, the temperature and the random number with a uniform distribution within $ (-0.5,0.5) $. 

\subsection{Parallelization of LEbc in ESPResSo++}\label{sec:parallel}

\begin{figure}[H]
	\begin{center}
		\includegraphics[trim=0px 0px 0px
		0px,clip,width=0.95\linewidth]{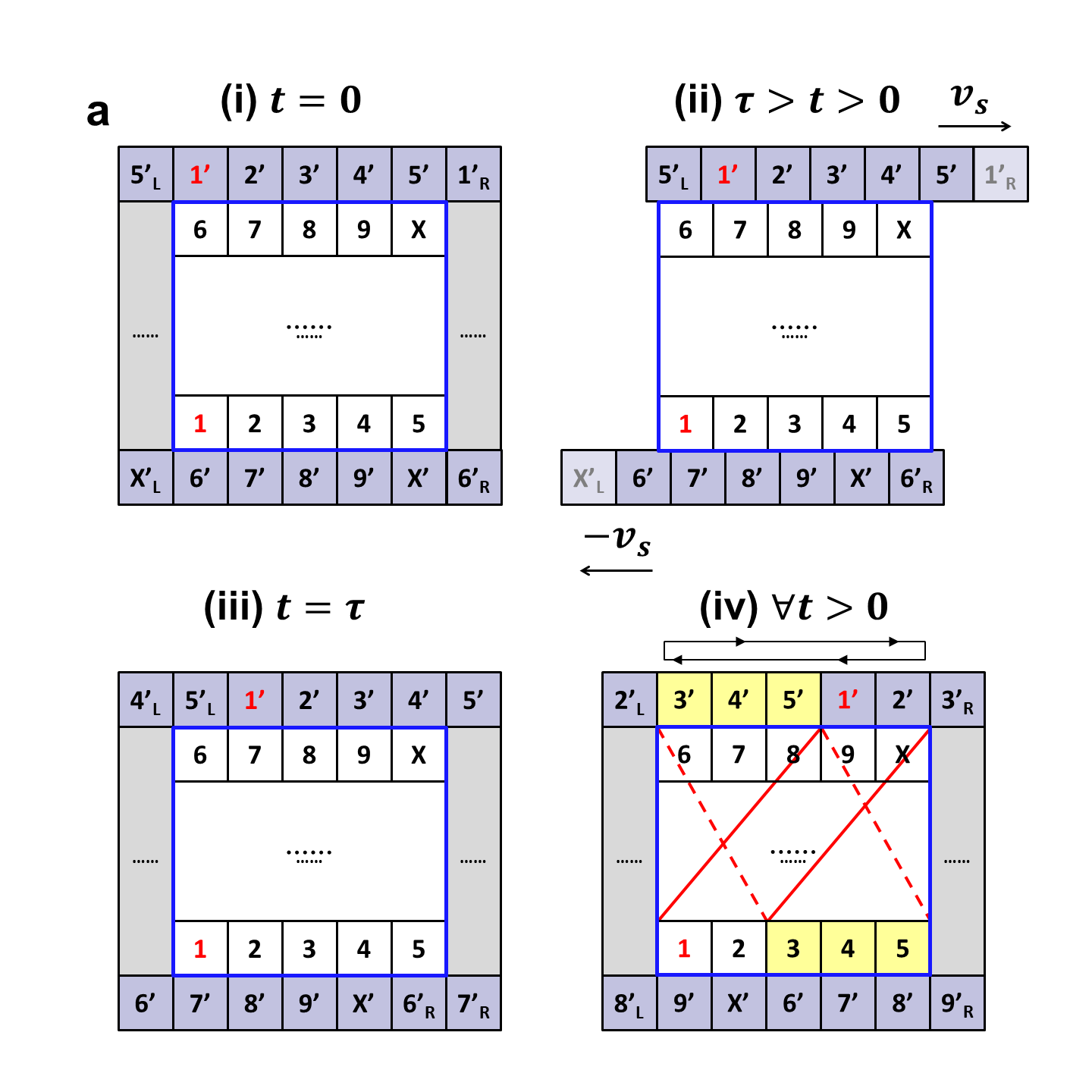}
	\end{center}
a)
\end{figure}

\begin{figure}[H]
	\begin{center}
		\includegraphics[trim=50px 100px 50px
		0px,clip,width=0.95\linewidth]{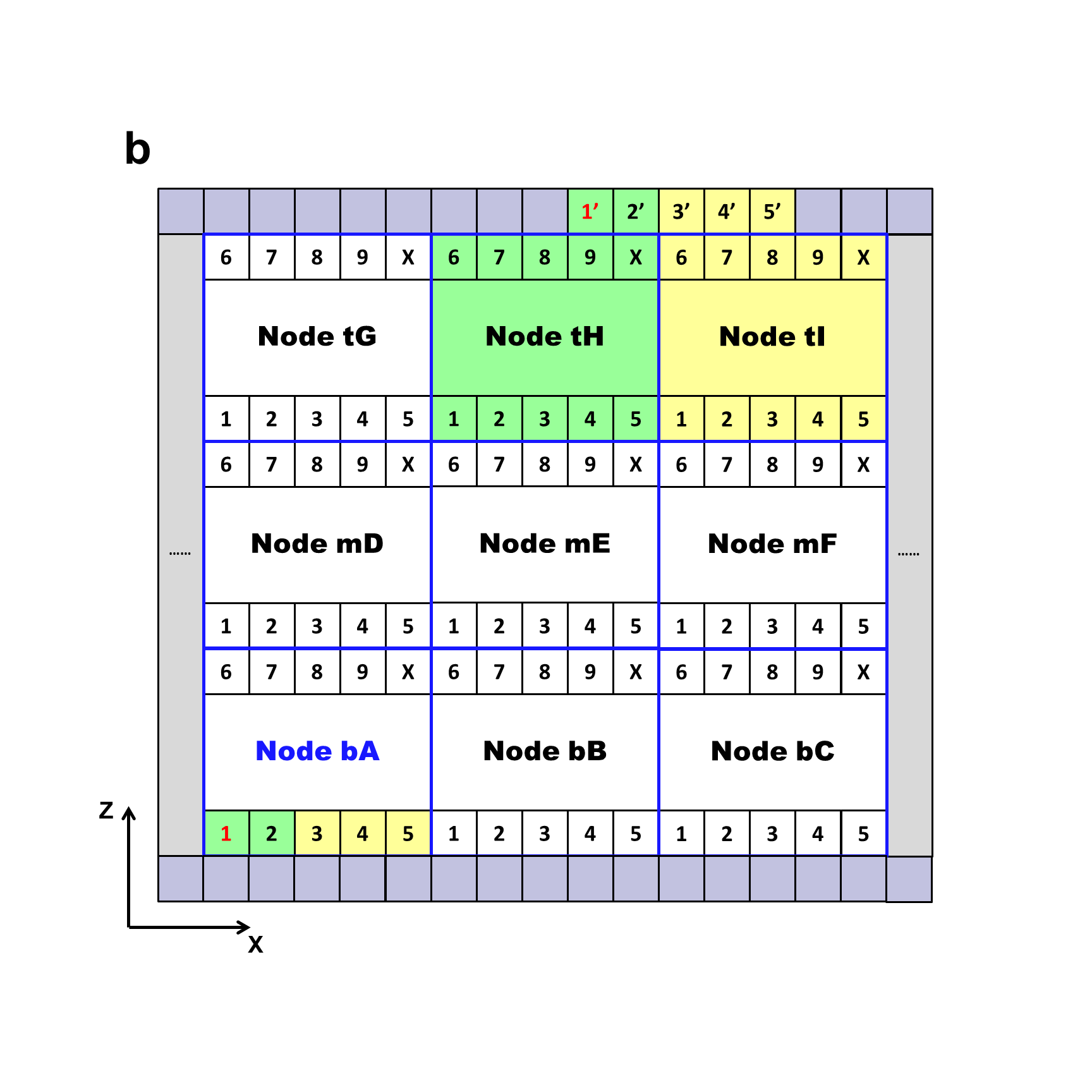}\\
		b)
		\caption{The pattern of cell communication within the domain decomposition for a) a single node grid (which stands for the serial computing) and b) 3$\times$3 node grids (parallel computing). The cells shift in the ghost layers (in gray) at the speed of $ \pm v_s(L_z/2)=\pm\dot{\gamma}L_z/2 $ when a shear flow starts. $ \tau $ is the duration for a ghost cell moving by a complete cell grid, which is given by $ \tau=2l_\text{cell}/(\dot{\gamma}L_z) $ and $ l_\text{cell} $ is the size of each cell in the $x-$direction.
		The ghost cell also shift iteratively (panel a-iv), and thus the connections of neighbor cells for data communication change dynamically.
		The panel b shows the generalization of dynamic cell communication with the presence of multiple node grids in parallel computing.
		In this pattern, node grids are assigned to their respective MPI ranks.
		When the ghost cells in the top ghost layer shift to the position as shown in the panel b, the current data communication (operated by the MPI communicator) occurs between Node \textbf{bA} and \textbf{tH} via Cell \gcell{1} and \gcell{2} (light green), and between Node \textbf{bA} and \textbf{tI} via Cell \gcell{3}$ - $ and \gcell{5} (light yellow).
		}
		\label{fig:decomp}
	\end{center}
\end{figure}
%
Consider an equilibrium MD simulation in a cuboid system with the domain decomposition which subdivides the simulation box into N$_x\times$N$_y\times$N$_z$ cell grids.
If the pairwise short-range interactions are present, only particle pairs with in a distance cutoff $r_c$ are included for force calculations.
However, a boundary condition should be taken into account while these distances are measured.
If two particles of a particle pair are close to two opposite boundaries respectively, an image transformation is required for computing their distance.
Thus, within the domain decomposition ghost cells are introduced (gray in Fig. \ref{fig:decomp}a-i)) and directly (up- or downright) mapped from real cells within the central domain (the box with blue frame lines).
For each cell mapping, the particle information (masses, positions and properties etc.) is copied from the real cell to the ghost cell. Once the force calculations are complete, in the ghost cell updated forces are sent backwards to the real cell for the next step of velocity integration.
In \epp, ghost cells (or layers) are built in an order of $ x $-,  $ y $- and $ z $-directions, and for simplicity the $ y $-direction is omitted in Figure \ref{fig:decomp}.
At $ t=0 $, the ghost cells in the ghost layers are indexed as \gcell{1}, \gcell{2}, $ \cdots $ and so on, corresponding to the real cells from the central domain.
For ghost cells at the corners, there is no direct mapping from real cells but instead particle information is copied from ghost cells which are created from previous iteration of the cell communication (e.g., \textbf{1}\sups{\prime}$\rightarrow$\rcell{1} or \textbf{5}\sups{\prime}$\rightarrow$\lcell{5}).
If the simulation starts without a shear, the starting network between ghost and real cells for data transmission remains unchanged during all simulation time.
According to the 26-neighbor rule in rebuilding neighbor lists\cite{DD-PLIMPTON1995}, for example, Cell \textbf{7} collects particle information from its ghost neighbors, namely Cell \textbf{1}\sups{\prime}, \textbf{2}\sups{\prime} and \textbf{3}\sups{\prime}. 
And Cell \textbf{7} is also responsible for receiving particles leaving from Cell \textbf{1}, \textbf{2} and \textbf{3}, via the corresponding ghost cells as virtual transitions.
For a shear simulation, however, all ghost cells start moving to the left (for cells at the bottom) and right (for cells at the top, see Fig. \ref{fig:decomp}a-ii).
Once those ghost cells have moved by one cell size ($l_\text{cell}$); for example, Cell \textbf{1}\sups{\prime} has been shifted by one complete cell grid, at time $ t=\tau=2l_\text{cell}/(\dot{\gamma}L_z) $, and posits perfectly above Cell \textbf{7}.
Thereby, the new neighbor cells (\lcell{5}, \textbf{1}\sups{\prime} \& \textbf{2}\sups{\prime}) are re-assigned to the real cell \textbf{7} and particle pairs are updated in its neighbor list (Fig. \ref{fig:decomp}a-iii).
Meanwhile, ghost cells at the corners are updated correspondingly (i.e., \rcell{1} and \lcell{X} replaced by \lcell{4} and \rcell{7}).
These new cell links will last for the next time $ \tau $ until the ghost cells shift to the next grid positions.
More accurately, Cell \textbf{7} starts receiving Cell \lcell{5} (replacing Cell \gcell{3}), \textbf{1}\sups{\prime} and \textbf{2}\sups{\prime} as neighbors at $ t=\tau/2 $ as Cell \textbf{1}\sups{\prime} has moved by a half grid and being the closest neighbor cell to Cell \textbf{7} among all ghost cells.
These new neighbor cells are valid until $t=3/2\tau$.
The update of neighbor cells is also an iterative process, see in Fig. \ref{fig:decomp}a-iv. With the rightward shift of in the top ghost layer, the ghost cell \gcell{1} finally leaves from the rightmost boundary and reappears to its original position at $t=0$ as in Fig. \ref{fig:decomp}a-i.
That is also a result of that the image box completes a move by a full length of $L_x$.
\\
\\
Figure \ref{fig:decomp}b generalizes the pattern of the cell communication  in the domain decomposition when many computing processors are present and all data communication is operated by the MPI (Messaging Passing Interface) communication in a shear flow simulation.
Similar to the simulations in serial computing, the data communication relies on ghost cells in the shear planes.
The difference in parallel computing, however, is that ghost cells are not external or attached to the entire central simulation box but bind to each node domain.
The data communication contains mainly two parts: 1) sending particle information from real cells to ghost cells (\textbf{r2g}) and 2) receiving results of force calculations from ghost cells and back to real cells (\textbf{g2r}).
For the equilibrium simulation, the \rg~communication follows, for the example of the target node \textbf{bA} and in $z-$direction, the sequence of nodes $\textbf{mD}\rightarrow\textbf{bA}\rightarrow \textbf{tG}$ (see the first column of nodes in Fig. \ref{fig:decomp}b and the right arrow represents the direction for sending particle information).
The particle information from Node \textbf{bA} is transmitted to Node \textbf{tG} via ghost cells above the boundary.
The \gr~communication, on the other hand, undergoes in reversed direction - nodes $\textbf{tG}\rightarrow\textbf{bA}\rightarrow\textbf{mD}$  and all node connections are left unchanged until the end of EQMD simulation.
Similar to the cell communication in serial computing for the shear flow simulations, the connection between node grids is also dynamic for every node at the top (labeled as node  \textbf{tX} and $ \textbf{X}=\{\textbf{G},\textbf{H},\textbf{I}\cdots\} $, see Fig. \ref{fig:decomp}b) and bottom (\textbf{bY}, $ \textbf{Y}=\{\textbf{A},\textbf{B},\textbf{C}\cdots\} $).
Following the example pattern in Fig. \ref{fig:decomp}b, dual nodes \textbf{tH}~and \textbf{tI} receive the particle information from and return the forces back to Node \textbf{bA}, via cells \{\gcell{1}, \gcell{2}\} and cells \{\gcell{3}, \gcell{4}, \gcell{5}\} respectively.
In the reversed $z-$direction, the communication of $\textbf{tH}\rightarrow\textbf{bA}$ and $\textbf{tI}\rightarrow\textbf{bA}$ is also taking effects at the mean time and respectively via 2 and 3 ghost cells as well.
Compared to the linear route of the communication $\textbf{mD}\leftrightarrow\textbf{bA}\leftrightarrow\textbf{tG}$ in EQMD, the communication in NEMD requires the MPI communicator to request extra communication between MPI ranks which satisfies dual routes between the node grids (one from the top shear plane; the other from the bottom) like $\textbf{bA}\Leftrightarrow \textbf{tH/tI}$.
For the communication with the internal node grids (\textbf{mZ}, $ \textbf{Z}=\{\textbf{D},\textbf{E},\textbf{F}\cdots\} $), the linear route stays unchanged (e.g., $\textbf{mD}\leftrightarrow\textbf{bA}$) during the shear flow simulation.

\subsection{Simulation Details and Analysis}
All MD simulations were carried out by using the \epp~MD software package\cite{EPP-Halverson2013,EPP-Guzman2019} and Lennard-Jones fluids and Kremer-Grest (KG) polymer melts\cite{KG-Grest1986,KG-Kremer1990,POLYM-Shang2017} were chosen as model systems for validating the implementation of the current work.
Before the shear flow simulations start, all systems were (heated and) equilibrated under equilibrium simulations.
The \lj~fluid contains a total of $ N=2000 $ particles in a cubic box.
The density of the system is set to $ \rho=0.844 $ and the \lj~potential is given by
\begin{equation}
\begin{aligned} 
V(r) = 4 \epsilon \left[ \left( \frac{\sigma}{r} \right)^{12} -
\left( \frac{\sigma}{r} \right)^{6} \right],
\end{aligned}
\label{eq:lj}
\end{equation}
where $ r $ is the distance between paired particles.
Other physical parameters are set unitless to $ \epsilon=\sigma=m=k_\text{B}=1$.
The short-range interactions have a cutoff at $r_c=2.5\sigma$.
Simulations were run using both the Langevin thermostat and the dissipative particle dynamics method.
For the Langevin thermostat, the friction coefficients of $\xi=1$ and $100$ are chosen to represent low and high viscosities of fluids; 
For the DPD thermostat, the friction constants of 5 and 25 are chosen.
Temperatures were set to $ T=0.5,1.0 $ and $ 1.5 $ for studying different liquid behaviors.
\\
\\
The KG polymer melts were modeled in forms of linear chains of beads\cite{KG-Grest1986,KG-Kremer1990}.
The total number of polymer beads are fixed to $ N=4000 $ and the density of the system is $ \rho=0.84 $.
The numbers of monomers for each polymer chain (namely the length of a chain) were selected to $m=20, 50$ and 100, respectively.
All polymer melts use the shifted \lj~potential 
\begin{equation}
\begin{aligned} 
V_\text{LJ}(r) = 4 \epsilon \left[ \Big( \frac{\sigma}{r} \Big)^{12} -\Big( \frac{\sigma}{r_c} \Big)^{12} -
\Big( \frac{\sigma}{r} \Big)^{6} +\Big( \frac{\sigma}{r_c} \Big)^{6} \right],
\end{aligned}
\label{eq:lj-shift}
\end{equation}
at the short cutoff of $r_c=2^{1/6}\sigma$\cite{LJ-WCA}.
Within a polymer chain, the bonded interactions between two adjacent beads are modeled using the FENE (finitely extensible nonlinear elastic\cite{KG-Kremer1990}) potential

\begin{equation}
V_\text{FENE}(r) = \left\{ \begin{array}{lr}
	-\dfrac{1}{2}kr_\text{max}^2ln\left[1-\left(\dfrac{r}{r_\text{max}}\right)^2\right], & \text{if }r<r_\text{max}, \\
	+\infty, & \text{if }  r\geq r_\text{max}
	\end{array}
\color{white}\right\}
,
\label{eq:fene}
\end{equation}
\\
where $k=30$ refers to the force constant and $r_\text{max}=1.5\sigma$ is the maximal bond length.
The angular term is presented in a form of cosine potential, which writes as

\begin{equation}
\begin{aligned} 
V_{cos}(r) = k_a\left(1-\phi/\phi_0\right)
\end{aligned}
\label{eq:angle}
\end{equation}
with the force constant $ k_a=1.5 $ and the equilibrium $ \phi_0=180^\circ $.
\\
\\
Combining Eq. \ref{eq:lj-shift}-Eq. \ref{eq:angle}, the overall total potential for a KG melt system is written as
\begin{equation}
\begin{aligned} 
E_\text{tot} = \sum V_\text{LJ}(r)+\sum V_\text{FENE}(r)+\sum V_{cos}(r)
\end{aligned}
\label{eq:kg-tot}
\end{equation}
for all particle pairs and triples under corresponding selection conditions.
\\
\\
In addition, the shear viscosity was computed for both model systems.
The general non-Newtonian shear viscosity $ \eta $ can be obtained, at finite shear rates, by calculating
\begin{equation}
\begin{aligned} 
\eta=\frac{\left<\varsigma_{xy}\right>}{\dot{\gamma}},
\end{aligned}
\label{eq:viscosity}
\end{equation}
where $ \varsigma_{xy} $ denotes as the off-diagonal component of the shear stress tensor \bfit{$\varsigma$}.
\bfit{$\varsigma$} is given by the Irving-Kirkwood equation\cite{Irving1950}

\begin{equation}
\begin{aligned} 
\varsigma=-\dfrac{1}{V}\left[\sum_{i} m\left(v_i-v_{s,i}\right)\otimes\left(v_i-v_{s,i}\right)+\sum_{i}\sum_{j (j>i)}r_{ij}\otimes F_{ij}(r_{ij})\right],
\end{aligned}
\label{eq:tensor}
\end{equation}
where $ V $ is the volume of the simulation box, $ v_{s,i} $ is the instantaneous shear velocity for particle $ i $ (which can be obtained from Eq. \ref{eq:shear-rate}), $ F_{ij} $ stands for the conservative force between particles $ i  $ and $ j $ and the $ \otimes $ symbol denotes to the dyadic product.
\section{Results and Discussion}\label{sec:result}
\subsection{Lennards-Jones fluids}
\begin{figure}[H]
	\begin{center}
		\includegraphics[trim=0px 0px 0px
		0px,clip,width=0.95\linewidth]{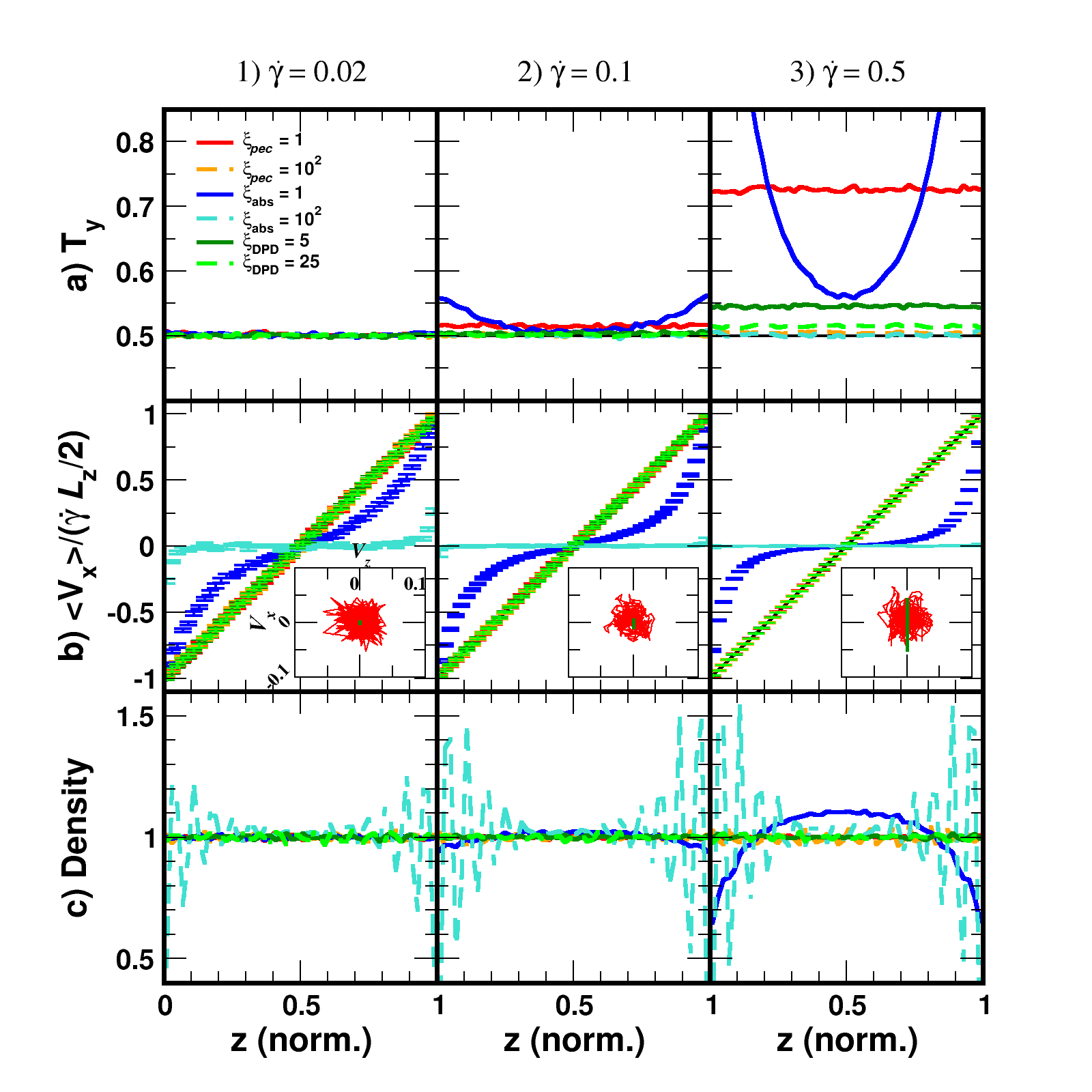}\\
		i) $T=0.5$
	\end{center}
\end{figure}
\begin{figure}[H]
\begin{center}
	\includegraphics[trim=0px 0px 0px
	50px,clip,width=0.95\linewidth]{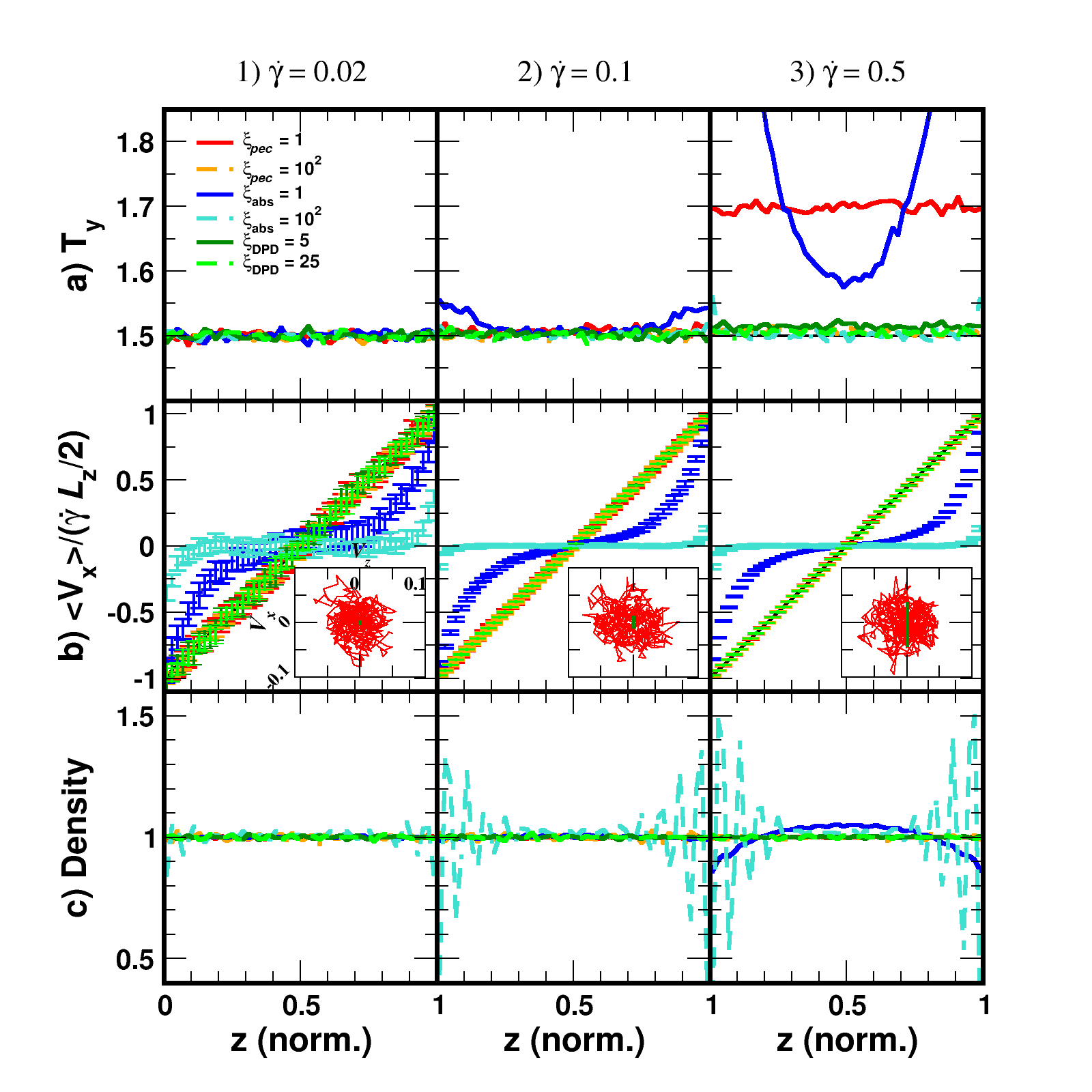}\\
	ii) $T=1.5$
	\caption{Profiles of physical observables (\textbf{a} - temperature; \textbf{b} - velocity; \textbf{c} - density) for MD simulations with shear rates at $\dot{\gamma}=0.02,0.1$ and 0.5 and temperatures at i) $T=0.5$ and ii) $T=1.5$.
		All data are presented as a function of $z-$coordinate (normalized to $0\sim1$) and obtained by averaging 50 independent MD trajectories (the same number of averaging also applies in the following of this section, if not specified).
		For each MD trajectory, only the last 20\% of the MD steps are used.
		$\xi_{pec}$ and $\xi_{\text{abs}}$ represent the friction coefficients in the Langevin thermostat. The dragging forces with $\xi_{pec}$ in the Langevin thermostat is only applied upon \textit{peculiar} velocity of particles without the shear contribution, whereas the absolute velocities are thermostatted with $\xi_\text{abs}$.
		In the panel \textbf{b} charts are with error bars which are manually amplified by $\times250\%$.
		Additionally, the insets in panel \textbf{b} report the momentum trajectories (projected on the $x-z$ plane) of the centers of mass from example MD runs using the Langevin ($\xi_{pec}$, red) and DPD (green) thermostats and with other corresponding input parameters.
	}
	\label{fig:profile}
\end{center}
\end{figure}

Figure \ref{fig:profile} presents the profiles of the configurational temperature $ T_y $ (along $y-$direction), the one-dimensional velocity $ v_x $ and the density $ \rho $ as a function of the $z-$coordinate.
The aim is to discuss the liquid behavior in different layers along the gradient direction of shear.
The MD simulations were run with selected temperatures and different ways of thermalization ($ \xi_{pec} $, $\xi_{\text{abs}}$ and $\xi_\text{DPD}$).
At the low shear rate ($ \dot{\gamma}=0.02 $), all simulations are well thermalized to their target temperatures.
For the velocity $ v_x $, however, different behaviors are found.
For simulations run with the Langevin thermostat ($ \xi_{pec} $) and the DPD thermostat, the linear velocity profiles are reproduced as expected. For the $\xi_{\text{abs}}$ thermalization, however, the non-linear development by layers are seen in all selected temperatures (Fig. \ref{fig:profile}i/ii-b1).
Especially for the simulations with $\xi_{\text{abs}}=100$ and at $ T=0.5 $, in more than $ 80\% $ of the layers particles are strongly stuck despite of the presence of the shear flow.
\\
\\
Similar observations were also found in the density profile (Fig. \ref{fig:profile}i/ii-c1).
The uniformed distribution of the density was seen in both of the $ \xi_{pec} $ and $\xi_\text{DPD}$ thermalization, which suggests the agreement of homogeneity between the simulation systems and real physics.
With $\xi_{\text{abs}}=100$, results fail to show such homogeneity and a layering behavior is found for layers close to the boundaries. 
That implies it is not necessary to keep a linear velocity profile even if the temperature is well maintained by the thermostat.
Such layering effects are even stronger for higher shear rates ($ \dot{\gamma}=0.1 $ and 0.5).
These unrealistic inhomogeneities in the both velocity and density profiles for $\xi_{\text{abs}}=1$ indicate that a conventional Langevin thermostat is not good enough to simulate a homogeneous system with a steady shear flow and reproduce some physical observables correctly.
Hence, simulations with the $\xi_{\text{abs}}$ thermalization are not further discussed in this paper.
\\
\\
As a refinement to the conventional Langevin thermostat, the $ \xi_{pec} $ thermalization applies the Langevin dynamics upon the peculiar velocity.
With $ \xi_{pec} $, both linear velocity profile and flat density profile are recovered by any combination of friction coefficients, shear rates and temperatures, as well as results from DPD simulations which are usually considered as a reference.
When the simulations of a higher shear rate ($ \dot{\gamma}=0.5 $) were investigated, however, the refined Langevin thermostat does not show a good thermalization especially if a weaker viscous friction and/or a lower temperature are present.
For simulations with $ \xi_{pec}=1 $ and $T=0.5$, the temperature profile shows a more than $+40\%$ up from the target temperature (Fig. \ref{fig:profile}i-a3).
Only a higher temperature or a extremely strong friction reduces or even diminishes the temperature deviation.
Interestingly, the DPD thermostat also somewhat fails to thermalize the system at the highest shear rate by $+10\%$ up from the target temperature.
That could indicate that it is necessary to control the shear flow in speed in order to fully dissipate extra kinetic energies for simulation introduced by fast shears.
At last, the momentum conservation was compared between the $ \xi_{pec}$ and $\xi_\text{DPD}$ thermalization.
Since all particle masses are equal to 1, the velocities of the center of mass (CoM) were instead computed.
As shown in the inset graphs in Fig. \ref{fig:profile}i/ii-b, the velocity trajectories of CoM are given.
For each simulation input, one representative trajectory is presented.
The DPD thermostat is well known for its momentum conserving feature.
At the low shear rate ($\dot{\gamma}=0.02$), the momentum conservation is successfully reproduced in all dimensions.
At the high shear rate ($\dot{\gamma}=0.5$), however, the influences from the shear flows are seen that the momentum conservation is not rigorous especially on the shear direction.
Nevertheless, the momentum of CoM in the $y-$ and $z-$directions is conserved, which makes the velocity trajectory line-segment like, see in the insets of Fig. \ref{fig:profile}i/ii-b3.
For the Langevin thermostat, the velocity trajectories fluctuate around the origin during the MD simulations.
The majorities of fluctuations are in the ranges of $ v_x$ or $v_z\in(-0.05,0.05) $ and $ \in(-0.1,0.1) $ at $ T=0.5 $ and $ T=1.5 $ respectively, which are much smaller than the averaged velocities (in one dimension) for a particle is given by $ v_\text{kin}=\sqrt{2T/m} $ from the kinetic energies at the given temperature ($v_\text{kin}=1$ for $ T=0.5 $ and $v_\text{kin}=1.732$ for $ T=1.5 $).
In addition, it is not found that a high shear speed significantly influences the momentum of CoM in the simulations using the Langevin thermostat.
This could be attributed to the absence of the shear contribution  and the equi-distributed density along the gradient direction from the simulations using the $ \xi_{pec}$ thermalization.
\begin{figure}[H]
	\begin{center}
		\includegraphics[trim=0px 0px 0px 0px,clip,width=0.9\linewidth]{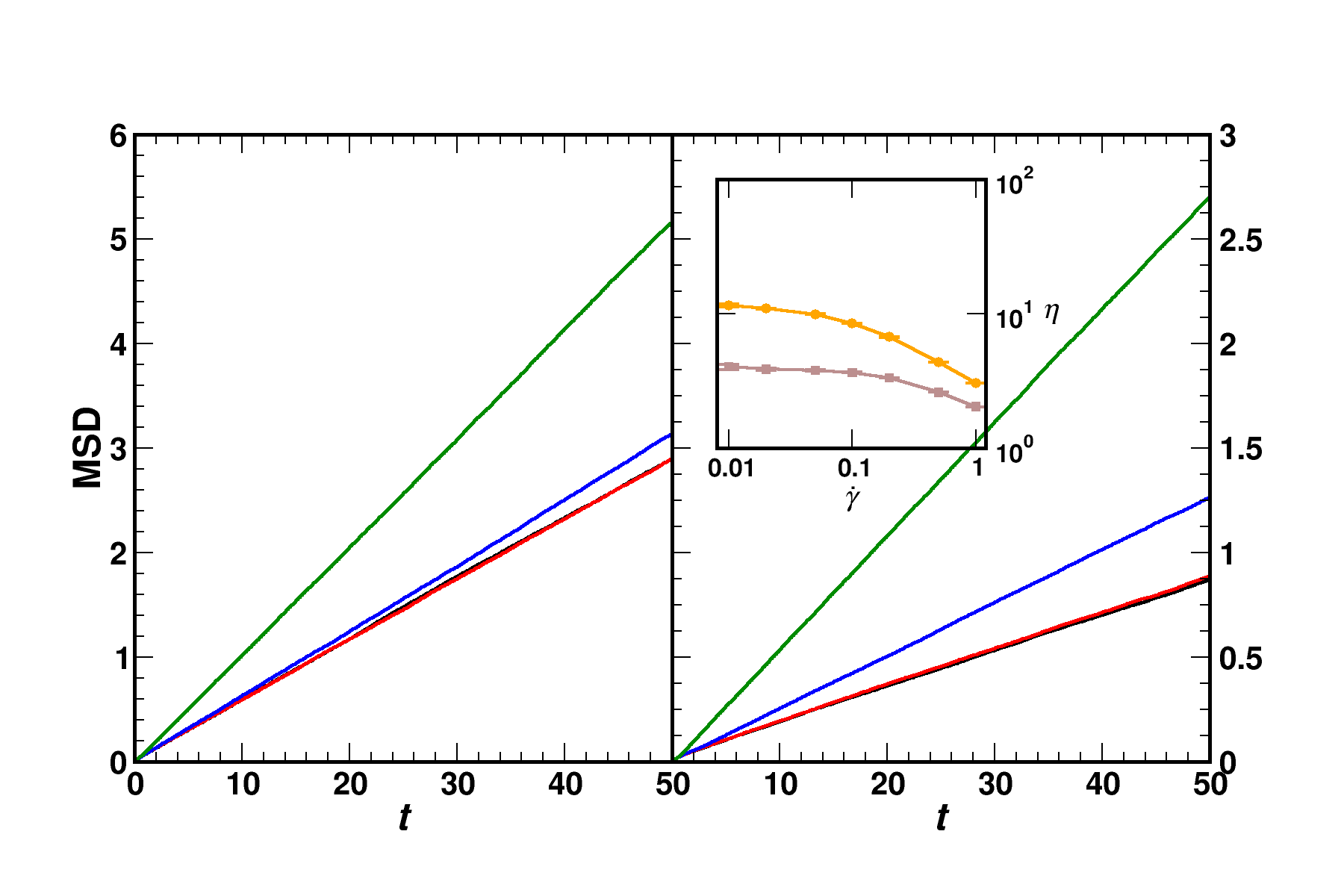}\\
		\caption{One dimensional mean squared displacement (MSD) of the \lj~fluids in the $ z- $direction under the shear flows.
		Simulations were run at $\dot{\gamma}=0.02$ (red), 0.1 (blue) and 0.5 (green) and with the Langevin thermostat (left : $\xi_\text{DPD}=5$; right: $\xi_\text{DPD}=25$) at $ T=1.0 $.
		Data were compared to reference MSD from the equilibrium simulation at the same temperature (see black solid lines).
		The inset shows the shear viscosity as a function of shear rates, ranged from $\dot{\gamma}=0.01$ to $\dot{\gamma}=1$ and data were computed for both $\xi_\text{DPD}=5$ (brown) and $\xi_\text{DPD}=25$ (orange).
		}
		\label{fig:msd}
	\end{center}
\end{figure}
\noindent
To investigate the system diffusivity of the \lj~fluids under LEbc, we first compared mean squared displacement of \lj~particles along the non-shear direction (the $z-$dimensions) at selected shear rates.
Figure \ref{fig:msd} shows such results from shear flow simulations with the DPD thermostat at different friction coefficients.
The self-diffusion constant (SDC) can be obtained by calculating $ D=\dfrac{1}{2}\left<x(t)^2\right>/t $, namely the half of the slope.
The overall SDCs of the sheared \lj~fluid have around twofold increases in the simulations with  ($\xi_\text{DPD}=5$) for every selected shear rate, compared to those with $\xi_\text{DPD}=25$.
The reference particle displacement is also computed from the EQMD simulations (see black solid lines).
If a steady shear starts, the diffusivity rises with the increase of the shear rate which even applies to non-shear directions.
We also investigated the relationship between the shear viscosity and the shear rate.
For both $\xi_\text{DPD}=5$ and $\xi_\text{DPD}=25$, that is consistent with the measurement on viscosity which monotonically decreases with the increase of $\dot{\gamma}$, see the inset of Figure \ref{fig:msd}.

\subsection{Kremer-Grest Polymer Melts}
\begin{figure}[H]
	\begin{center}
		\includegraphics[trim=0px 0px 0px
		0px,clip,width=0.8\linewidth]{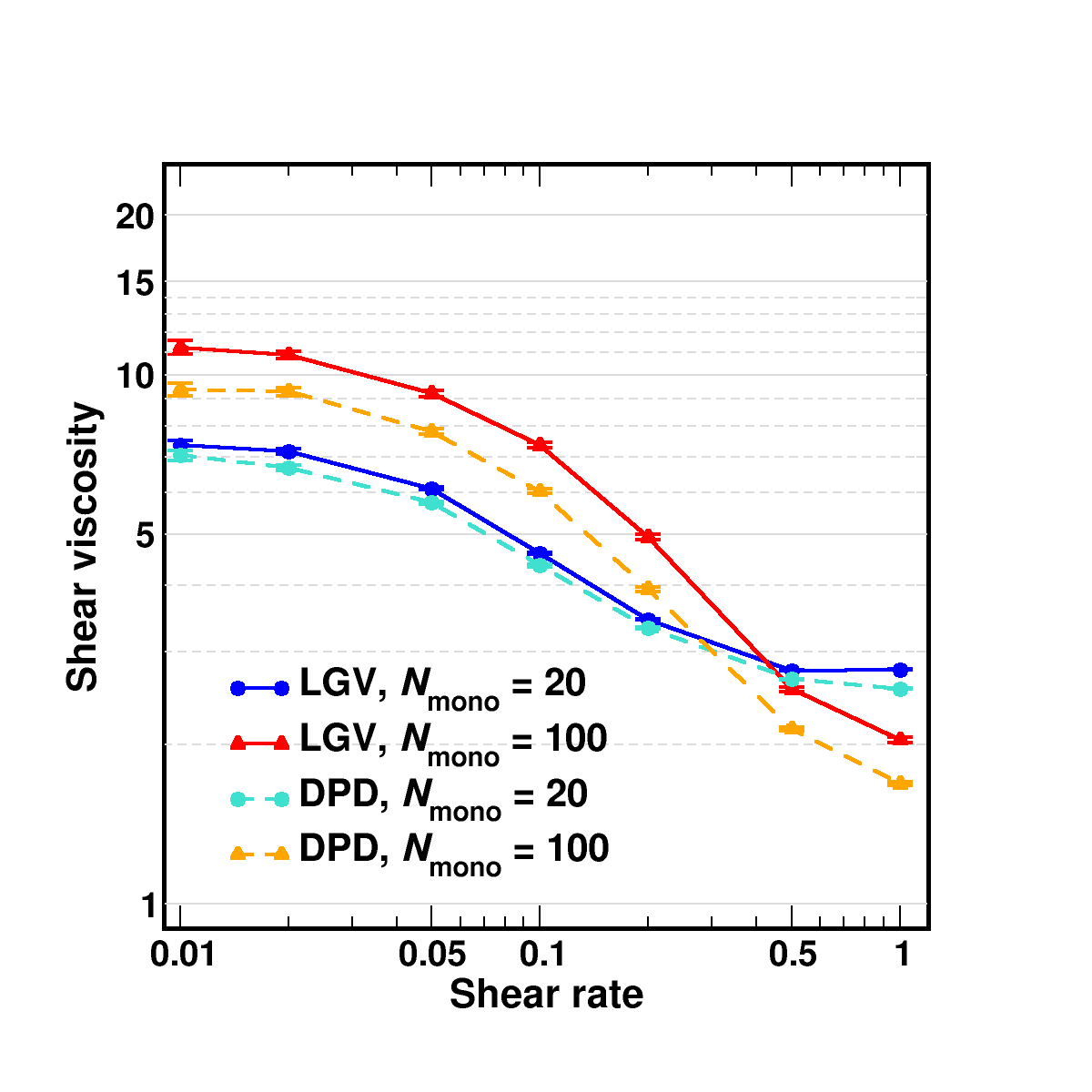}\\
		\caption{Shear viscosity for polymer melts ($N_\text{mono}=20$ and 100 per chain) in the shear flow simulation using Langevin thermostat and with Lees-Edwards boundary condition. Data were reported for shear rates in a range of $\dot{\gamma}=0.01\sim1.0$.
		}
		\label{fig:visco}
	\end{center}
\end{figure}
The most common type of behavior to a shear flow model of polymer melts is the shear thinning, where the fluid viscosity decreases with an increasing shear rate.
Figure \ref{fig:visco} shows the shear viscosity for simulations using both the Langevin and DPD thermostats.
$ N_\text{mono} $ represents the length of monomers for each chain in the polymer melt systems and the shear rate $\dot{\gamma}$ ranges from 0.01 to 1.
In general, all results show similar behaviors with respect to the shear thinning.
At the low shear rates ($\dot{\gamma}<0.02$), the polymer melts show more Newtonian behaviors with a constant value of shear viscosity, which finally converges to the plateau of zero-shear viscosity  as $ \lim\{\dot{\gamma}\rightarrow0\} $.
The low shear rate region applies to simulations of both lengths of polymer chains.
At the region of a critical shear rate ($0.02<\dot{\gamma}<0.5$ for $ N_\text{mono}=20 $), the shear viscosity drops enormously and usually a power law model can be applied for numeric fitting.
For $\dot{\gamma}>0.5$, the region of the highest shear rates, the decrease of the shear viscosity slows down which means the viscosity becomes less sensitive to higher shear stress when polymer chains are fully disentangled and aligned.
In this region, a second viscosity plateau can be observed (or called the infinite shear viscosity plateau) or a weaker shear thinning behavior forms which can be fitted to a Sisko model\cite{Sisko}.
For the longer polymer chains, the second viscosity plateau has yet to be seen within the region in Figure \ref{fig:visco} but still the trending of slowdown can be seen in $0.5<\dot{\gamma}<1$.

\subsection{Performance Benchmark}
\begin{figure}[H]
	\begin{center}
		\includegraphics[trim=0px 20px 0px
		50px,clip,width=0.8\linewidth]{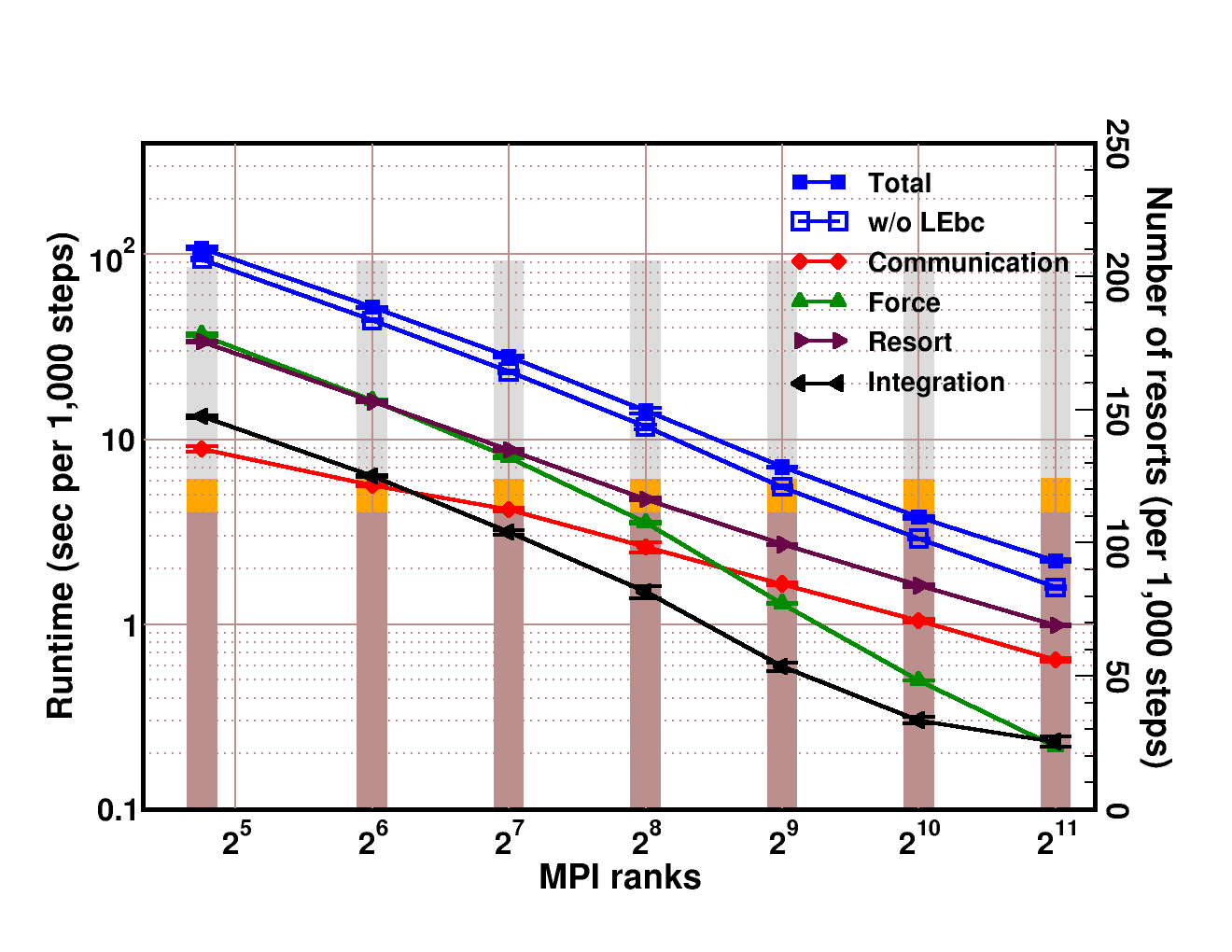}\\
		\caption{Total runtime of parallel computing with different MPI ranks for shear simulations of ring polymer melts (with a total of $N=2,560,000$ beads and 200 beads per chain). The time costs from the modules of communication (red), force calculations (green), resort (maroon) and the integration (black) are presented as the contributions into the total runtimes. The numbers of resorts are given in bar forms. All results are collected and averaged from 20 independent MD runs and error bars are also present (though very trivial as shown)
		}
		\label{fig:benchmark}
	\end{center}
\end{figure}
In Sec. \ref{sec:parallel}, we have discussed the adaption of the \lebc~for parallel computing and the MPI communication with dual routes for exchanging particle and force information between the node grids.
This specialized real-to-ghost and ghost-to-real communication, however, only occurs between the node grids from the top panel of the shear plane within the simulation box and those from the bottom panel (let us call them \emph{special} nodes).
For other node grids located in the internal part of the central box, their communication to those \emph{special} node grids and to each other of themselves obeys the standard scheme of the domain decomposition as if the equilibrium simulations are run.
\epp~offers two modes to proceed the communication scheme for a target node grid.
For internal-internal and internal-\emph{special} node pairs, the first mode is applied which uses the standard pattern of cell communication; For \emph{special}-\emph{special} node pairs, the second mode is switched on and the LEbc adapted communication is used.
However, that limits the minimum number of node grids in the $z-$direction ($ N_z$).
If $ N_z=2 $, the neighbor nodes (at the top and at the bottom) do not only communicate cross the $xy-$boundary planes but also communicate to each other internally within the central box.
That requires the activation for both modes, which is not supported by the current data layouts.
Therefore, the LEbc based domain decomposition is required to set $ N_z\ge3 $.
In the current version of \epp, the assignment of the numbers of node grids (with given MPI ranks) from the domain decomposition is fixed to $ N_x\ge N_y\ge N_z $.
Thus, for running a simulation with LEbc there is a minimum requirement of at least 27 MPI ranks for parallel computing.
But this would be trivial, as a problem, since we focus on many-particle systems ($N>1,000,000$) in future works.
\\
\begin{table}
\caption{Comparison of the total runtime  (per 1,000 steps) between the shear simulation with LEbc and non-shear simulation (without LEbc).}
\begin{center}
\begin{tabular}{ c c c c c c  }
\hline
  \multirow{2}{*}{$N_\text{core}$} & \multicolumn{3}{c}{Node and cell grids}& \multicolumn{2}{c}{Total runtime (s)} \\
\cline{2-6}
    & \multicolumn{3}{c}{$ \left[N_x,N_y,N_z\right]\times\left(n_x,n_y,n_z\right) $} & LEbc & w/o LEbc  \\
\hline
27    & \multicolumn{3}{c}{$\left[3,3,3\right]\times\left(21,21,21\right)$} & 107.8 & 94.2  \\ 
64    & \multicolumn{3}{c}{$\left[4,4,4\right]\times\left(16,16,16\right)$} & 51.6 & 43.9   \\ 
128   & \multicolumn{3}{c}{$\left[8,4,4\right]\times\left(8,16,16\right)$}  & 27.9 & 23.2   \\ 
256   & \multicolumn{3}{c}{$\left[8,8,4\right]\times\left(8,8,16\right)$}   & 14.3 & 11.7   \\ 
512   & \multicolumn{3}{c}{$\left[8,8,8\right]\times\left(8,8,8\right)$}    & 7.1 & 5.5     \\ 
1024  & \multicolumn{3}{c}{$\left[16,8,8\right]\times\left(4,8,8\right)$}   & 3.8 & 2.9     \\ 
2048  & \multicolumn{3}{c}{$\left[16,16,8\right]\times\left(4,4,8\right)$}  & 2.2 & 1.6     \\ 
\hline
\hline
  \multirow{2}{*}{$N_\text{core}$} & \multicolumn{5}{c}{Performance difference (w/o LEbc as reference)}   \\
\cline{2-6}
 & Total & Communication & Force & Resort & Integration\\
 \hline
27    & $+14.5\%$ & $-12.2\%$& $-2.6\%$ & $+101\%$ & $-4.3\%$    \\
64    & $+17.6\%$ & $+1.6\%$ & $-1.6\%$ & $+104\%$ & $-3.0\%$    \\
128   & $+20.4\%$ & $+4.6\%$ & $+0.8\%$ & $+106\%$ & $-0.9\%$    \\
256   & $+20.8\%$ & $+1.4\%$ & $+2.5\%$ & $+107\%$ & $-5.9\%$    \\
512   & $+26.6\%$ & $+1.9\%$ & $+2.7\%$ & $+109\%$ & $+3.6\%$    \\
1024  & $+30.1\%$ & $+0.9\%$ & $+1.9\%$ & $+108\%$ & $+2.7\%$    \\
2048  & $+39.8\%$ & $+7.6\%$ & $+0.8\%$ & $+122\%$ & $+16.7\%$   \\
\hline
\end{tabular}
\end{center}
\label{tab:runtime}
\end{table}
\\
All benchmarks were performed on the MOGON II supercomputer cluster from the data center of Johannes Gutenberg University Mainz.
Up to 64 nodes were used for performing the shear flow simulation for a huge sized polymer melt system (a total of $ N=2,560,000 $ beads).
Each node contains two 16-core Intel Skylake processors (Xeon Gold 6130, 2.10GHz) and nodes are networked with 100 Gbps Omni-Path.
The Intel compiler (v2018.03) is used for compiling the \epp~software package and the Langevin thermostat was used for the benchmark simulations.
As shown in Figure \ref{fig:benchmark} and Table \ref{tab:runtime}, the benchmark simulation starts from one node parallel computing with $ N_\text{core}=27 $ which creates $ 3\times3\times3 $ node grids by the domain decomposition.
At $ N_\text{core}=27 $, the average total runtime is $t=107.8$ seconds per 1,000 MD steps for shear simulation with LEbc, compared to 94.2 seconds for its EQMD counterpart (see the hollow square).
Thus, the overall performance differs in $14.5\%$ and
Figure \ref{fig:benchmark} also presents the contributions of time costs.
At $ N_\text{core}=27 $ the parts of force calculation (including both bonded and non-bonded interactions) and resorts take the majority of the contributions among all computing procedures.
With the increase of MPI ranks, only the resort part remains as the main contributor from $ N_\text{core}\ge256 $.
For the force calculation, some super-scaling effects were found, due to the data caching when the simulation is highly parallelized, and its place in time consumption is the overtaken by the communication.
\\
\\
As mentioned in Sec. \ref{sec:parallel}, using LEbc requires a more complex pattern in data communication. 
Hence, it is also of our interests in further analyzing factors which lead to the slowdown in the overall runtime between the shear flow simulation and the EQMD simulation with zero-shear.
Table \ref{tab:runtime} also compares such difference in time costs for all contributions between EQMD and NEMD simulations.
Apparently, the slowdown in the shear flow simulation is predominantly attributed to the resort of which the time costs are increased by $+100\%$ $\sim$ $+120\%$.
This is mainly due to more frequent calls on the resort function during the shear simulation.
In Fig. \ref{fig:benchmark}, the numbers of resorts are reported for both EQMD and NEMD simulations.
For the EQMD simulation, the resort is called by 111 times per 1,000 MD steps, which is also found insensitive to the number of MPI ranks.
For NEMD with shear flow, the number rises up to 205 times per 1,000 MD steps.
This agrees with the factor of two in the difference of the time cost for the resort part.
It is also worth to note that some resorts are forcibly activated when a ghost cell completes a shift move (as discussed for Fig. \ref{fig:decomp}a-ii).
That is because every shift move requires to rebuild a new connection of neighbor cells.
Such enforced resorts occur by 12.7 times per 1,000 MD steps, shown as short orange bars in Fig. \ref{fig:benchmark} and approximately $\sim$6$\%$ out of the total number of resorts.
By taking the number of resorts from EQMD simulations as the background frequency of resorts and excluding the contribution from enforced resorts, the shear contribution to the resorts can be then obtained from the the non-equilibrium dynamics (see the gray bars).
For the shear simulation using LEbc, the dual routes create more complexity during the MPI communication and it becomes suspicious on whether extra overheads or load imbalance are introduced which might potentially hinder the speed of the computation.
Such concern, however, is clarified when the difference of performance for the communication is compared (between EQMD and NEMD) and no more than $ +6\% $ deviation is found for all MPI ranks.
That could be explained by the fact that the dual routes only change the source and destination MPI processors from the one-to-one (MPI rank) mode to the one-to-two mode.
The total amount of the data communication (if counted in paired cells) for each \emph{special} node is unchanged since the global homogeneity of the system remains stable under the steady shear.
In the strategy of \emph{columnar} domain decomposition, introduced by Bindgen \emph{et al.} into the ESPResSoMD package, all ghost cells are considered the neighbor cells to the real cells at the boundaries of the shear plane.
That consequently leads to an all-to-all mapping between the \emph{special} nodes when the MPI communicator is called (i.e., the communication between Node \textbf{bA} and nodes \textbf{tG}, \textbf{tH} and \textbf{tI} is always on during the entire simulation).
By this, rebuilding connection of neighbour cells (which corresponds to the enforced resorts in this paper) can be avoided but at costs of extra communication (including potential load imbalance) and bringing more work loads to the force calculator.
Potentially, it may also introduce an extra scaling factor ($\bfit{o}\left(\sqrt[3]{N_\text{core}}\right)$) for the shear simulation which makes it more difficult to take advantage from parallel computing when simulating larger sized systems ($N>100,000$).
For the LEbc implementation in this paper, we proved that the additional cost given by the enforced resorts are trivial, compared to that of the background resorts from the dynamics.
By identifying the connection of neighbor cells on the fly, the intensity of MPI communication is under control for each \emph{special} node by limiting paired nodes to a fixed number and the total input sent to the force calculation is minimized without missing any pair under $r_\text{cutoff}$.
Using the dual routes guarantees that the overall performance of the shear simulation is comparable to its EQMD counterpart in the \epp~package and the linearity of scaling is, to a great extent, preserved using many computing processors.

\section{Conclusions and outlooks}\label{sec:summary}
In this paper, we have presented the implementation of the \lebc~in the \epp~MD software package and the detailed design of its parallelization.
We hope the current work can be helpful and inspiring also for future codes or implementations in other open source MD software.
\\
\\
We employed our LEbc implementation in non-equilibrium MD simulations using both Langevin and DPD thermostats.
We first investigated the \lj~fluids and found that using the conventional Langevin thermostat fails to reproduce the linear profiles of velocities.
Moreover, the strong layering effects for the density profiles were found for the regions close to the boundaries which are perpendicular to the gradient of the shear.
Instead, a modified Langevin thermostat, which acts on the \emph{peculiar} velocities, was used and certain physical observables were recovered.
Results also agree with the simulations using the DPD thermostat at the low and moderate shear rates.
\\
\\
We also simulated Kremer-Grest polymer melts and determined the shear thinning region between $\dot{\gamma}=0.02$ and $\dot{\gamma}=0.5$ for short chains polymer melts.
To measure the parallel performance for the current implementation, we used up to 32 supercomputer nodes for simulating a huge polymer melt system with more than $2.5$ million particles.
Results show the overall speedup from $N_\text{core}=64$ to $N_\text{core}=1024$ is around $\times1200\%$ (and the scaling efficiency is ca. $75\%$).
The major overhead created for many processors parallelization can be attributed to more frequent resorts during the MD simulations which is inevitable with the presence of a shear flow.
\\
\\
For the future work, we first aim to exploit the future performances in parallel computing.
Possible directions could include 1) the optimization of resort management, 2) the integration of the current LEbc code into the new load-balance optimization from the recent \epp~development\cite{EPP-Vance2021} and 3) further optimization of the data communication and the data layouts which are intensively used in the domain decomposition.
Second, we are also expecting to extend the current LEbc implementation for more generalized application cases. For example, we are interested in finding more commonly used thermostats (i.e, the Bussi-Danadio-Parrinello thermostat\cite{BDP-Bussi2007}) and integrating them with the LEbc implementation.
We also notice that there will be increasing needs for performing shear flow simulations for systems which require long-range interactions.
In a first step, Ewald \cite{Ewald-Wheeler1997} is now an available method in \epp~for simulations with LEbc.
Further applications and code optimization will be presented in future publications.
\clearpage

\section*{Acknowledgments}
This work has been supported by the Deutsche Forschungsgemeinschaft (DFG) through the Collaborative Research Center Transregio TRR 146 Multiscale Simulation Methods for Soft Matter Systems. Parts of this research were conducted using the supercomputer Mogon II and advisory services offered by the Zentrum für Datenverarbeitung at the Johannes Gutenberg University Mainz (hpc.uni-mainz.de).

\bibliography{library}

\providecommand{\latin}[1]{#1}
\makeatletter
\providecommand{\doi}
  {\begingroup\let\do\@makeother\dospecials
  \catcode`\{=1 \catcode`\}=2 \doi@aux}
\providecommand{\doi@aux}[1]{\endgroup\texttt{#1}}
\makeatother
\providecommand*\mcitethebibliography{\thebibliography}
\csname @ifundefined\endcsname{endmcitethebibliography}
  {\let\endmcitethebibliography\endthebibliography}{}
\begin{mcitethebibliography}{38}
\providecommand*\natexlab[1]{#1}
\providecommand*\mciteSetBstSublistMode[1]{}
\providecommand*\mciteSetBstMaxWidthForm[2]{}
\providecommand*\mciteBstWouldAddEndPuncttrue
  {\def\EndOfBibitem{\unskip.}}
\providecommand*\mciteBstWouldAddEndPunctfalse
  {\let\EndOfBibitem\relax}
\providecommand*\mciteSetBstMidEndSepPunct[3]{}
\providecommand*\mciteSetBstSublistLabelBeginEnd[3]{}
\providecommand*\EndOfBibitem{}
\mciteSetBstSublistMode{f}
\mciteSetBstMaxWidthForm{subitem}{(\alph{mcitesubitemcount})}
\mciteSetBstSublistLabelBeginEnd
  {\mcitemaxwidthsubitemform\space}
  {\relax}
  {\relax}

\bibitem[Ruiz-Franco \latin{et~al.}(2018)Ruiz-Franco, Rovigatti, and
  Zaccarelli]{LJ-Ruiz-Franco2018}
Ruiz-Franco,~J.; Rovigatti,~L.; Zaccarelli,~E. On the effect of the thermostat
  in non-equilibrium molecular dynamics simulations. \emph{Eur. Phys. J. E}
  \textbf{2018}, \emph{41}, 80\relax
\mciteBstWouldAddEndPuncttrue
\mciteSetBstMidEndSepPunct{\mcitedefaultmidpunct}
{\mcitedefaultendpunct}{\mcitedefaultseppunct}\relax
\EndOfBibitem
\bibitem[Leach(2001)]{MM-Leach2001}
Leach,~A. \emph{Molecular Modelling: Principles and Applications}; Pearson
  Education; Prentice Hall, 2001\relax
\mciteBstWouldAddEndPuncttrue
\mciteSetBstMidEndSepPunct{\mcitedefaultmidpunct}
{\mcitedefaultendpunct}{\mcitedefaultseppunct}\relax
\EndOfBibitem
\bibitem[MacKerell~Jr. \latin{et~al.}()MacKerell~Jr., Brooks~III, Nilsson,
  Roux, Won, and Karplus]{CHARMMFF}
MacKerell~Jr.,~A.; Brooks~III,~C.; Nilsson,~L.; Roux,~B.; Won,~Y.; Karplus,~M.
  \emph{CHARMM: The Energy Function and Its Parameterization with an Overview
  of the Program}; The Encyclopedia of Computational Chemistry\relax
\mciteBstWouldAddEndPuncttrue
\mciteSetBstMidEndSepPunct{\mcitedefaultmidpunct}
{\mcitedefaultendpunct}{\mcitedefaultseppunct}\relax
\EndOfBibitem
\bibitem[Vanommeslaeghe \latin{et~al.}(2009)Vanommeslaeghe, Hatcher, Acharya,
  Kundu, Zhong, Shim, Darian, Guvench, Lopes, Vorobyov, and Mackerell]{CGenFF}
Vanommeslaeghe,~K.; Hatcher,~E.; Acharya,~C.; Kundu,~S.; Zhong,~S.; Shim,~J.;
  Darian,~E.; Guvench,~O.; Lopes,~P.; Vorobyov,~I.; Mackerell,~A.~D. CHARMM
  general force field: A force field for drug-like molecules compatible with
  the CHARMM all-atom additive biological force fields. \emph{J. Comp. Chem.}
  \textbf{2009}, \emph{31}, 671--690\relax
\mciteBstWouldAddEndPuncttrue
\mciteSetBstMidEndSepPunct{\mcitedefaultmidpunct}
{\mcitedefaultendpunct}{\mcitedefaultseppunct}\relax
\EndOfBibitem
\bibitem[Xu and Meuwly(2019)Xu, and Meuwly]{MMPT-Xu2019}
Xu,~Z.-H.; Meuwly,~M. Multistate Reactive Molecular Dynamics Simulations of
  Proton Diffusion in Water Clusters and in the Bulk. \emph{J. Phys. Chem. B}
  \textbf{2019}, \emph{123}, 9846--9861\relax
\mciteBstWouldAddEndPuncttrue
\mciteSetBstMidEndSepPunct{\mcitedefaultmidpunct}
{\mcitedefaultendpunct}{\mcitedefaultseppunct}\relax
\EndOfBibitem
\bibitem[Xu and Meuwly(2017)Xu, and Meuwly]{MMPT-Oxa-XuMeuwly2017}
Xu,~Z.-H.; Meuwly,~M. Vibrational Spectroscopy and Proton Transfer Dynamics in
  Protonated Oxalate. \emph{J. Phys. Chem. A} \textbf{2017}, \emph{121},
  5389--5398\relax
\mciteBstWouldAddEndPuncttrue
\mciteSetBstMidEndSepPunct{\mcitedefaultmidpunct}
{\mcitedefaultendpunct}{\mcitedefaultseppunct}\relax
\EndOfBibitem
\bibitem[Mackeprang \latin{et~al.}(2016)Mackeprang, Xu, Maroun, Meuwly, and
  Kjaergaard]{MMPT-Mackeprang-2016}
Mackeprang,~K.; Xu,~Z.-H.; Maroun,~Z.; Meuwly,~M.; Kjaergaard,~H.~G.
  Spectroscopy and Dynamics of Double Proton Transfer in Formic Acid Dimer.
  \emph{Phys. Chem. Chem. Phys.} \textbf{2016}, \emph{18}, 24654--24662\relax
\mciteBstWouldAddEndPuncttrue
\mciteSetBstMidEndSepPunct{\mcitedefaultmidpunct}
{\mcitedefaultendpunct}{\mcitedefaultseppunct}\relax
\EndOfBibitem
\bibitem[Frenkel and Smit(2002)Frenkel, and Smit]{Book_Understanding2002}
Frenkel,~D.; Smit,~B. In \emph{Understanding Molecular Simulation (Second
  Edition)}, second edition ed.; Frenkel,~D., Smit,~B., Eds.; Academic Press:
  San Diego, 2002; pp 63--107\relax
\mciteBstWouldAddEndPuncttrue
\mciteSetBstMidEndSepPunct{\mcitedefaultmidpunct}
{\mcitedefaultendpunct}{\mcitedefaultseppunct}\relax
\EndOfBibitem
\bibitem[Evans and Morriss(1990)Evans, and Morriss]{Book_EVANS1990_NE_LIQUIDS}
Evans,~D.~J.; Morriss,~G.~P. \emph{Statistical Mechanics of Nonequilibrium
  Liquids}; Academic Press, 1990; pp 1--10\relax
\mciteBstWouldAddEndPuncttrue
\mciteSetBstMidEndSepPunct{\mcitedefaultmidpunct}
{\mcitedefaultendpunct}{\mcitedefaultseppunct}\relax
\EndOfBibitem
\bibitem[Hoover and Hoover(2005)Hoover, and Hoover]{NEMD-Hoover2004}
Hoover,~W.~G.; Hoover,~C.~G. {Nonequilibrium molecular dynamics}.
  \emph{Condens. Matter Phys.} \textbf{2005}, \emph{8}, 247--260\relax
\mciteBstWouldAddEndPuncttrue
\mciteSetBstMidEndSepPunct{\mcitedefaultmidpunct}
{\mcitedefaultendpunct}{\mcitedefaultseppunct}\relax
\EndOfBibitem
\bibitem[Sarman \latin{et~al.}(1998)Sarman, Evans, and
  Cummings]{NEMD-Sarman1998}
Sarman,~S.; Evans,~D.~J.; Cummings,~P. Recent developments in non-Newtonian
  molecular dynamics. \emph{Phys. Rep.} \textbf{1998}, \emph{305}, 1--92\relax
\mciteBstWouldAddEndPuncttrue
\mciteSetBstMidEndSepPunct{\mcitedefaultmidpunct}
{\mcitedefaultendpunct}{\mcitedefaultseppunct}\relax
\EndOfBibitem
\bibitem[Todd and Daivis(2007)Todd, and Daivis]{SLLOD_Todd2007}
Todd,~B.~D.; Daivis,~P.~J. Homogeneous non-equilibrium molecular dynamics
  simulations of viscous flow: techniques and applications. \emph{Mol.
  Simulat.} \textbf{2007}, \emph{33}, 189--229\relax
\mciteBstWouldAddEndPuncttrue
\mciteSetBstMidEndSepPunct{\mcitedefaultmidpunct}
{\mcitedefaultendpunct}{\mcitedefaultseppunct}\relax
\EndOfBibitem
\bibitem[Khare \latin{et~al.}(1997)Khare, {de Pablo}, and Yethiraj]{KharedPY97}
Khare,~R.; {de Pablo},~J.; Yethiraj,~A. Molecular simulation and continuum
  mechanics study of simple fluids in non-isothermal planar couette flows.
  \emph{The Journal of Chemical Physics} \textbf{1997}, \emph{107},
  2589--2596\relax
\mciteBstWouldAddEndPuncttrue
\mciteSetBstMidEndSepPunct{\mcitedefaultmidpunct}
{\mcitedefaultendpunct}{\mcitedefaultseppunct}\relax
\EndOfBibitem
\bibitem[Rastogi \latin{et~al.}(1996)Rastogi, Wagner, and
  Lustig]{LEBC-Rastogi1996-Wall}
Rastogi,~S.~R.; Wagner,~N.~J.; Lustig,~S.~R. Rheology, self‐diffusion, and
  microstructure of charged colloids under simple shear by massively parallel
  nonequilibrium Brownian dynamics. \emph{J. Chem. Phys.} \textbf{1996},
  \emph{104}, 9234--9248\relax
\mciteBstWouldAddEndPuncttrue
\mciteSetBstMidEndSepPunct{\mcitedefaultmidpunct}
{\mcitedefaultendpunct}{\mcitedefaultseppunct}\relax
\EndOfBibitem
\bibitem[Tuckerman \latin{et~al.}(1997)Tuckerman, Mundy, Balasubramanian, and
  Klein]{SLLOD_Tuckerman1997}
Tuckerman,~M.~E.; Mundy,~C.~J.; Balasubramanian,~S.; Klein,~M.~L. Modified
  nonequilibrium molecular dynamics for fluid flows with energy conservation.
  \emph{J. Chem. Phys.} \textbf{1997}, \emph{106}, 5615--5621\relax
\mciteBstWouldAddEndPuncttrue
\mciteSetBstMidEndSepPunct{\mcitedefaultmidpunct}
{\mcitedefaultendpunct}{\mcitedefaultseppunct}\relax
\EndOfBibitem
\bibitem[Petravic and Evans(1998)Petravic, and Evans]{SLLOD_Petravic1998}
Petravic,~J.; Evans,~D.~J. Approach to the non-equilibrium time-periodic state
  in a ‘steady’ shear flow model. \emph{Mol. Phys.} \textbf{1998},
  \emph{95}, 219--231\relax
\mciteBstWouldAddEndPuncttrue
\mciteSetBstMidEndSepPunct{\mcitedefaultmidpunct}
{\mcitedefaultendpunct}{\mcitedefaultseppunct}\relax
\EndOfBibitem
\bibitem[Lees and Edwards(1972)Lees, and Edwards]{LEBC_Lees1972}
Lees,~A.~W.; Edwards,~S.~F. The computer study of transport processes under
  extreme conditions. \emph{J. Phys. C: Solid State Phys.} \textbf{1972},
  \emph{5}, 1921--1928\relax
\mciteBstWouldAddEndPuncttrue
\mciteSetBstMidEndSepPunct{\mcitedefaultmidpunct}
{\mcitedefaultendpunct}{\mcitedefaultseppunct}\relax
\EndOfBibitem
\bibitem[Wagner and Pagonabarraga(2002)Wagner, and
  Pagonabarraga]{LEBC-Wagner2002-LB}
Wagner,~A.~J.; Pagonabarraga,~I. Lees-Edwards Boundary Conditions for Lattice
  Boltzmann. \emph{J. Stat. Phys.} \textbf{2002}, \emph{107}, 521--537\relax
\mciteBstWouldAddEndPuncttrue
\mciteSetBstMidEndSepPunct{\mcitedefaultmidpunct}
{\mcitedefaultendpunct}{\mcitedefaultseppunct}\relax
\EndOfBibitem
\bibitem[Rastogi and Wagner(1996)Rastogi, and Wagner]{LEBC-Rastogi1996}
Rastogi,~S.~R.; Wagner,~N.~J. A parallel algorithm for Lees-Edwards boundary
  conditions. \emph{Parallel Comput.} \textbf{1996}, \emph{22}, 895--901\relax
\mciteBstWouldAddEndPuncttrue
\mciteSetBstMidEndSepPunct{\mcitedefaultmidpunct}
{\mcitedefaultendpunct}{\mcitedefaultseppunct}\relax
\EndOfBibitem
\bibitem[Bindgen \latin{et~al.}(2021)Bindgen, Weik, Weeber, Koos, and
  de~Buyl]{LEBC-Bindgen2021}
Bindgen,~S.; Weik,~F.; Weeber,~R.; Koos,~E.; de~Buyl,~P. Lees–Edwards
  boundary conditions for translation invariant shear flow: Implementation and
  transport properties. \emph{Physics of Fluids} \textbf{2021}, \emph{33},
  083615\relax
\mciteBstWouldAddEndPuncttrue
\mciteSetBstMidEndSepPunct{\mcitedefaultmidpunct}
{\mcitedefaultendpunct}{\mcitedefaultseppunct}\relax
\EndOfBibitem
\bibitem[Halverson \latin{et~al.}(2013)Halverson, Brandes, Lenz, Arnold, Bevc,
  Starchenko, Kremer, Stuehn, and Reith]{EPP-Halverson2013}
Halverson,~J.~D.; Brandes,~T.; Lenz,~O.; Arnold,~A.; Bevc,~S.; Starchenko,~V.;
  Kremer,~K.; Stuehn,~T.; Reith,~D. ESPResSo++: A modern multiscale simulation
  package for soft matter systems. \emph{Comput. Phys. Commun.} \textbf{2013},
  \emph{184}, 1129--1149\relax
\mciteBstWouldAddEndPuncttrue
\mciteSetBstMidEndSepPunct{\mcitedefaultmidpunct}
{\mcitedefaultendpunct}{\mcitedefaultseppunct}\relax
\EndOfBibitem
\bibitem[Guzman \latin{et~al.}(2019)Guzman, Tretyakov, Kobayashi, Fogarty,
  Kreis, Krajniak, Junghans, Kremer, and Stuehn]{EPP-Guzman2019}
Guzman,~H.~V.; Tretyakov,~N.; Kobayashi,~H.; Fogarty,~A.~C.; Kreis,~K.;
  Krajniak,~J.; Junghans,~C.; Kremer,~K.; Stuehn,~T. ESPResSo++ 2.0: Advanced
  methods for multiscale molecular simulation. \emph{Comput. Phys. Commun.}
  \textbf{2019}, \emph{238}, 66--76\relax
\mciteBstWouldAddEndPuncttrue
\mciteSetBstMidEndSepPunct{\mcitedefaultmidpunct}
{\mcitedefaultendpunct}{\mcitedefaultseppunct}\relax
\EndOfBibitem
\bibitem[Grest and Kremer(1986)Grest, and Kremer]{KG-Grest1986}
Grest,~G.~S.; Kremer,~K. Molecular dynamics simulation for polymers in the
  presence of a heat bath. \emph{Phys. Rev. A} \textbf{1986}, \emph{33},
  3628--3631\relax
\mciteBstWouldAddEndPuncttrue
\mciteSetBstMidEndSepPunct{\mcitedefaultmidpunct}
{\mcitedefaultendpunct}{\mcitedefaultseppunct}\relax
\EndOfBibitem
\bibitem[Kremer and Grest(1990)Kremer, and Grest]{KG-Kremer1990}
Kremer,~K.; Grest,~G.~S. Dynamics of entangled linear polymer melts: A
  molecular‐dynamics simulation. \emph{J. Chem. Phys.} \textbf{1990},
  \emph{92}, 5057--5086\relax
\mciteBstWouldAddEndPuncttrue
\mciteSetBstMidEndSepPunct{\mcitedefaultmidpunct}
{\mcitedefaultendpunct}{\mcitedefaultseppunct}\relax
\EndOfBibitem
\bibitem[Soddemann \latin{et~al.}(2003)Soddemann, D\"unweg, and
  Kremer]{DPD_SoddemannKremer2003}
Soddemann,~T.; D\"unweg,~B.; Kremer,~K. Dissipative particle dynamics: A useful
  thermostat for equilibrium and nonequilibrium molecular dynamics simulations.
  \emph{Phys. Rev. E} \textbf{2003}, \emph{68}, 046702\relax
\mciteBstWouldAddEndPuncttrue
\mciteSetBstMidEndSepPunct{\mcitedefaultmidpunct}
{\mcitedefaultendpunct}{\mcitedefaultseppunct}\relax
\EndOfBibitem
\bibitem[Moshfegh and Jabbarzadeh(2015)Moshfegh, and
  Jabbarzadeh]{LEBC_Moshfegh2015}
Moshfegh,~A.; Jabbarzadeh,~A. Modified Lees–Edwards boundary condition for
  dissipative particle dynamics: hydrodynamics and temperature at high shear
  rates. \emph{Mol. Simulat.} \textbf{2015}, \emph{41}, 1264--1277\relax
\mciteBstWouldAddEndPuncttrue
\mciteSetBstMidEndSepPunct{\mcitedefaultmidpunct}
{\mcitedefaultendpunct}{\mcitedefaultseppunct}\relax
\EndOfBibitem
\bibitem[Shang \latin{et~al.}(2017)Shang, Kröger, and
  Leimkuhler]{POLYM-Shang2017}
Shang,~X.; Kröger,~M.; Leimkuhler,~B. Assessing numerical methods for
  molecular and particle simulation. \emph{Soft Matter} \textbf{2017},
  \emph{13}, 8565--8578\relax
\mciteBstWouldAddEndPuncttrue
\mciteSetBstMidEndSepPunct{\mcitedefaultmidpunct}
{\mcitedefaultendpunct}{\mcitedefaultseppunct}\relax
\EndOfBibitem
\bibitem[Hoogerbrugge and Koelman(1992)Hoogerbrugge, and
  Koelman]{DPD-HoogerbruggeKoelman1992}
Hoogerbrugge,~P.~J.; Koelman,~J. M. V.~A. Simulating Microscopic Hydrodynamic
  Phenomena with Dissipative Particle Dynamics. \emph{Europhys. Lett.}
  \textbf{1992}, \emph{19}, 155--160\relax
\mciteBstWouldAddEndPuncttrue
\mciteSetBstMidEndSepPunct{\mcitedefaultmidpunct}
{\mcitedefaultendpunct}{\mcitedefaultseppunct}\relax
\EndOfBibitem
\bibitem[Espa{\~n}ol and Warren(1995)Espa{\~n}ol, and Warren]{DPD-Espanol1995}
Espa{\~n}ol,~P.; Warren,~P.~B. Statistical Mechanics of Dissipative Particle
  Dynamics. \emph{Europhys. Lett.} \textbf{1995}, \emph{30}, 191--196\relax
\mciteBstWouldAddEndPuncttrue
\mciteSetBstMidEndSepPunct{\mcitedefaultmidpunct}
{\mcitedefaultendpunct}{\mcitedefaultseppunct}\relax
\EndOfBibitem
\bibitem[Groot and Warren(1997)Groot, and Warren]{DPD-Groot1997}
Groot,~R.~D.; Warren,~P.~B. Dissipative particle dynamics: Bridging the gap
  between atomistic and mesoscopic simulation. \emph{J. Chem. Phys.}
  \textbf{1997}, \emph{107}, 4423--4435\relax
\mciteBstWouldAddEndPuncttrue
\mciteSetBstMidEndSepPunct{\mcitedefaultmidpunct}
{\mcitedefaultendpunct}{\mcitedefaultseppunct}\relax
\EndOfBibitem
\bibitem[Plimpton(1995)]{DD-PLIMPTON1995}
Plimpton,~S. Fast Parallel Algorithms for Short-Range Molecular Dynamics.
  \emph{J. Chem. Phys.} \textbf{1995}, \emph{117}, 1--19\relax
\mciteBstWouldAddEndPuncttrue
\mciteSetBstMidEndSepPunct{\mcitedefaultmidpunct}
{\mcitedefaultendpunct}{\mcitedefaultseppunct}\relax
\EndOfBibitem
\bibitem[Weeks \latin{et~al.}(1971)Weeks, Chandler, and Andersen]{LJ-WCA}
Weeks,~J.~D.; Chandler,~D.; Andersen,~H.~C. Role of Repulsive Forces in
  Determining the Equilibrium Structure of Simple Liquids. \emph{J. Chem.
  Phys.} \textbf{1971}, \emph{54}, 5237--5247\relax
\mciteBstWouldAddEndPuncttrue
\mciteSetBstMidEndSepPunct{\mcitedefaultmidpunct}
{\mcitedefaultendpunct}{\mcitedefaultseppunct}\relax
\EndOfBibitem
\bibitem[Irving and Kirkwood(1950)Irving, and Kirkwood]{Irving1950}
Irving,~J.~H.; Kirkwood,~J.~G. The Statistical Mechanical Theory of Transport
  Processes. IV. The Equations of Hydrodynamics. \emph{J. Chem. Phys.}
  \textbf{1950}, \emph{18}, 817--829\relax
\mciteBstWouldAddEndPuncttrue
\mciteSetBstMidEndSepPunct{\mcitedefaultmidpunct}
{\mcitedefaultendpunct}{\mcitedefaultseppunct}\relax
\EndOfBibitem
\bibitem[Deli\'c \latin{et~al.}(2002)Deli\'c, Marn, and Žunič]{Sisko}
Deli\'c,~M.; Marn,~J.; Žunič,~Z. The Sisko Model For Non-Newtonian Fluid Flow
  Using The Finite-Volume Method. \emph{Stroj. Vestn.-J. Mech. E.}
  \textbf{2002}, \emph{48}, 687--695\relax
\mciteBstWouldAddEndPuncttrue
\mciteSetBstMidEndSepPunct{\mcitedefaultmidpunct}
{\mcitedefaultendpunct}{\mcitedefaultseppunct}\relax
\EndOfBibitem
\bibitem[Vance \latin{et~al.}(2021)Vance, Xu, Stuehn, Tretyakov, Rammp, Eibl,
  and Brinkmann]{EPP-Vance2021}
Vance,~J.; Xu,~Z.-H.; Stuehn,~T.; Tretyakov,~N.; Rammp,~M.; Eibl,~S.;
  Brinkmann,~A. {Code modernization strategies for short-range non-bonded
  molecular dynamics simulations}. \emph{Preprint} \textbf{2021}, \relax
\mciteBstWouldAddEndPunctfalse
\mciteSetBstMidEndSepPunct{\mcitedefaultmidpunct}
{}{\mcitedefaultseppunct}\relax
\EndOfBibitem
\bibitem[Bussi \latin{et~al.}(2007)Bussi, Donadio, and
  Parrinello]{BDP-Bussi2007}
Bussi,~G.; Donadio,~D.; Parrinello,~M. Canonical sampling through velocity
  rescaling. \emph{J. Chem. Phys.} \textbf{2007}, \emph{126}, 014101\relax
\mciteBstWouldAddEndPuncttrue
\mciteSetBstMidEndSepPunct{\mcitedefaultmidpunct}
{\mcitedefaultendpunct}{\mcitedefaultseppunct}\relax
\EndOfBibitem
\bibitem[Wheeler \latin{et~al.}(1997)Wheeler, Fuller, and
  Rowley]{Ewald-Wheeler1997}
Wheeler,~D.~R.; Fuller,~N.~G.; Rowley,~R.~L. Non-equilibrium molecular dynamics
  simulation of the shear viscosity of liquid methanol: adaptation of the Ewald
  sum to Lees-Edwards boundary conditions. \emph{Mol. Phys.} \textbf{1997},
  \emph{92}, 55--62\relax
\mciteBstWouldAddEndPuncttrue
\mciteSetBstMidEndSepPunct{\mcitedefaultmidpunct}
{\mcitedefaultendpunct}{\mcitedefaultseppunct}\relax
\EndOfBibitem
\end{mcitethebibliography}

\end{document}